\documentclass[aip,jcp,superscriptaddress]{revtex4-1}
\usepackage{amsmath}
\usepackage{amssymb}
\usepackage{graphicx}
\usepackage{enumerate}
\usepackage{natbib}

\begin{document}

\title{Estimating time-correlation functions by sampling and unbiasing dynamically activated events.}

\author{Manuel Ath\`enes}
\affiliation{CEA, DEN, Service de Recherches de M\'etallurgie Physique, F-91191 Gif-sur-Yvette, France}

\author{Mihai-Cosmin Marinica}
\affiliation{CEA, DEN, Service de Recherches de M\'etallurgie Physique, F-91191 Gif-sur-Yvette, France}

\author {Thomas Jourdan}
\affiliation{CEA, DEN, Service de Recherches de M\'etallurgie Physique, F-91191 Gif-sur-Yvette, France}

\renewcommand{\baselinestretch}{1.6}\normalsize

\begin{abstract}

Transition path sampling is a rare-event method that estimates state-to-state time-correlation functions 
in many-body systems from samples of short trajectories. In this framework, it is proposed to bias the importance function using the lowest Jacobian eigenvalue moduli along the dynamical trajectory. A lowest eigenvalue modulus is related to the lowest eigenvalue of the Hessian matrix and is evaluated here using the Lanczos algorithm as in activation-relaxation techniques. This results in favoring the sampling of activated trajectories and enhancing the occurrence of the rare reactive trajectories of interest, those corresponding to transitions between locally stable states. 
Estimating the time-correlation functions involves unbiasing the sample of
simulated trajectories which is done using the multi-state Bennett acceptance
ratio (MBAR) method. To assess the performance of our  procedure, we compute the
time-correlation function associated with the migration of a vacancy in
$\alpha$-iron. The derivative of the estimated time-correlation function yields
a migration rate in agreement with the one given by transition
state theory. Besides, we show that the information relative to rejected
trajectories can be recycled within MBAR, resulting in a substantial speed-up. 
Unlike original transition path-sampling, our approach does not require computing the reversible work to 
confine the trajectory endpoints to a reactive state. 

\end{abstract}

\maketitle

\section{Introduction}

Rate constants of thermally activated processes are crucial parameters governing kinetic and transport properties of condensed matter. Estimating these rates directly using molecular dynamics or kinetic Monte Carlo typically remains a challenge as the probability of observing an event is very low at the simulation time-scale. In practice, a sufficient number of reactive events has to be collected so as to achieve high accuracy in the estimates of the rates. This task consumes considerable amounts of computational time or may even be prohibitive. 
%For instance, the migration of a vacancy in $\alpha$-Fe at 500K typically happens once every ten nanoseconds, while the usual time steps for molecular dynamics is of the order of the femtosecond. 
This limitation arises in the simulations of many physical, chemical or biological systems in which the structural evolution of clusters, molecules or proteins involves transitions between locally stable states occurring rarely during a computer simulation. 
To reduce the total simulation cost, several biased sampling methods have been developed in order to enhance the probability of observing the rare events.~\cite{DBCD1998,BCDG2002,VAMFW2007,AVW2009} Among them, transition path sampling~\cite{BCDG2002} (TPS) aims at estimating a time-correlation function between two indicator functions, $h_A$ and $h_B$, characterizing the reaction. The indicator function $h_{A(B)} (\mathbf{x})$ takes value $1$ if state $\mathbf{x} \in A(B)$ and value $0$ elsewhere, $A$ and $B$ being two basins traditionally referred to as reactant and product. In practice, the state-to-state time-correlation that is estimated in TPS is based on a path ensemble average rather than a time average. 
Denoting by $z = \{ \mathbf{x}(s) \}_{0 \leq s \leq \mathcal{T}}$ the trajectories of the path ensemble and of duration $\mathcal{T}$, TPS considers that the trajectory initial states $\mathbf{x}(0)$ are distributed according to $h_A \rho_A^\mathrm{eq}$, where $\rho_A^\mathrm{eq}$ denotes the equilibrium Boltzmann distribution normalized to one within basin $A$. 
The time-correlation function in TPS is the overall conditional probability that a trajectory from basin $A$ is in state $B$ at $t < \mathcal{T}$
\begin{equation}
\label{eq:correl}
C(t)= \int h_{A} [\mathbf{x}(0)] h_{B}[\mathbf{x}(t)] \mathrm{P}_{\mathrm{cond}} \left[ \mathcal{D}z | {\mathbf{x}\left(0\right)}\right] \rho_A^\mathrm{eq} \left[ d\mathbf{x}(0) \right].
\end{equation}
where $\mathrm{P_{cond}}( \mathcal{D}z | \mathbf{x}(0))$, the probability measure  of trajectory $z$, is conditioned on its initial state $\mathbf{x}(0)$. The probability measure of the initial states is $h_A\left[\mathbf{x}(0)\right] \rho_A^\mathrm{eq}\left[d\mathbf{x}(0)\right]$. 
This definition of the time-correlation function~\cite{chandler1978statistical,dellago2002transition} naturally applies to any stochastic model coupling the dynamics to a thermostat, such as the Langevin dynamics. It may also apply to a model in which the system is initially prepared in equilibrium and allowed to evolve in isolation (without exchanging heat with the environment) by a deterministic reversible dynamics. The latter kind of dynamical model is often studied as it well approximates situations where the times $t$ of relevance to rate processes are small compared with the time required for a significant heat exchange of the simulated system with the environment. 

%$= \exp \left\{ \beta \left[F_A - \mathcal{H} \left(\mathbf{x} \right) \right] \right\} d\mathbf{x}$, the free energy $F_A = - \beta^{-1} \ln \int_A \exp \left[ -\beta \mathcal{H} (\mathbf{x}) d \mathbf{x} \right]$ normalizes the distribution within $A$ and $\mathcal{H}$ is the system Hamiltonian.  

The conditional probability $C(t)$ plays an important role in the molecular simulations of rare events because it corresponds to the fraction of trajectories that are reactive. When fast molecular relaxations within basin $A$ and activated processes associated with transitions exiting basin $A$ occur on well-separated time scales, the derivative of the time-correlation function, ${dC(t)}/{dt}$, displays a plateau corresponding to the phenomenological rate constant $k_{A\rightarrow B}$. The phenomenological rate may be derived at macroscopic scale considering fluxes over populations or at atomic scale resorting to linear response theory.~\cite{chandler1987introduction,chandler1978statistical,frenkel2002understanding}
TPS estimates the time-correlation function $C(t)$ over a time interval $\left] 0 , \mathcal{T} \right]$ from samples of constrained trajectories. In practice (see Refs. \onlinecite{DBCD1998} and~\onlinecite{dellago1999calculation}), TPS factorizes $C(t)$ into the product of (i) a reduced time-correlation function $R_{t^\prime}(t)=C(t)/C(t^\prime)$, estimated by taking an average in the ensemble of transition paths and (ii) a constant factor $K_{t^\prime}=C(t^\prime)$, obtained from the reversible work required to progressively confine the trajectory endpoints beyond milestones marked out along the reaction up to the reactive state. 

We show herein that, for a thermally activated process, it is possible to
compute the time-correlation function $C(t)$ without confining the trajectory endpoint to a reactive state. 
Eigenvalues of the Jacobian matrix associated with the dynamics will be used to favor the sampling of active trajectories, 
those visiting negatively curved portions of the potential energy surface (i.e. saddle regions of instability between energy basins). 
This strategy involves post-processing the samples of active trajectories so as to simultaneously (i) correct for the bias and (ii) estimate the fractions of reactive segments yielding the entire correlation function $C(t)$, two requirements reflecting the two-step procedure of TPS. We show in the paper that the multi-state Bennett acceptance ratio method~\cite{shirts2008statistically} achieves this goal. The proposed method is referred to as SUNDAE as it consists in sampling and unbiasing dynamically activated events. 

This paper is organized as follows. The key ingredients of the methodological recipe are given in the five following sections.  
Section~\ref{sec:determinants} derives the conditional probabilities for generating trial trajectories. Section~\ref{sec:Verlet} describes the Verlet map. Section~\ref{sec:Eigenvalue-decomposition} decomposes the Jacobian matrix associated with the Verlet map. Section~\ref{sec:Transition-path-sampling} describes the path-sampling scheme and section~\ref{sec:Reaction-rate-constants-calculation} derives the various unbiasing estimators tested in this study. 
We show in particular that MBAR can be combined with waste-recycling,~\cite{frenkel:2004,frenkel:2006} a technique consisting in
including information from unselected proposals so as to reduce the statistical variance of the estimates.~\cite{delmas:2009}
SUNDAE is finally applied to the calculation of a rate constant associated with the migration of a vacancy in $\alpha$-iron (Sec.~\ref{Numerical results}). 

\section{Trajectory space and conditional probabilities\label{sec:determinants}}

Let $\{ q_i \}_{1 \leq i \leq 3N}$ and $\{ p_i \}_{1 \leq i \leq 3N}$ denote the canonical coordinates of the system. 
In the following we write the evolution equations in matricial form and denote vectors and matrices by lowercase and uppercase bold letters, respectively. We thus introduce the column vectors $\mathbf{q}$ and $\mathbf{p}$ for the positions and momenta. State $\mathbf{x}$ is also written as a column vector $(q_1,...,q_{3N},p_1,...,p_{3N})^{T}$, the upper $T$ denoting vector or matrix transposes. Vector $\mathbf{m}$ denotes a deterministic and invertible map defined on $E$, the phase space: $\mathbf{x} \in E \rightarrow \mathbf{m}(\mathbf{x}) \in E$. 
The inverse map is denoted by $\mathbf{m}^{-1}$. It is also a functional column vector. The Jacobian matrix of $\mathbf{m}$ is denoted by $(\mathbf{ \nabla m})^T$ : its components are $\left[ (\nabla \mathbf{m})^T \right]_{ij}= \nabla_j m_i$ where $\nabla_i$ represent the partial derivative $\tfrac{\partial}{\partial x_i}$. From the inverse function theorem, the Jacobian matrix of $\mathbf{m}^{-1}$ at $\mathbf{m}(\mathbf{x})$ is the matricial inverse of $\mathbf{ \nabla m}^T$ at $\mathbf{x}$ 
\[ \left(\mathbf{\nabla {m}^{-1}}\right)^T [\mathbf{m}(\mathbf{x})]= \left( (\mathbf{\nabla m})^T\right)^{-1}(\mathbf{x}).\]
Consequently, the Jacobian determinant of $\mathbf{m}^{-1}$ at $\mathbf{m}(\mathbf{x})$ is $\det[\mathbf{ \nabla m }]^{-1}$ at $\mathbf{x}$. Let $\delta_\mathbf{a}(d \mathbf{y})$ denote the Dirac measure at $\mathbf{a}$. 
The direct transition probability from $\mathbf{x}$ to $\mathbf{y}$ is
\begin{eqnarray}\label{eq:jacobian}
 \delta_\mathbf{x}\left(\mathbf{m}^{-1}(\mathbf{y})\right) = 
  \left|\det \left[ \mathbf{\nabla m} (\mathbf{x}) \right] \right| \delta_{\mathbf{m}(\mathbf{x})}\left(\mathbf{y}\right). 
\end{eqnarray} 
Similarly for the inverted map, the probability to transition from $\mathbf{y}$ to $\mathbf{x}$ (reverse transition) is 
\begin{equation}\label{eq:jacobian2}
\delta_\mathbf{y}\left[\mathbf{m}(\mathbf{x})\right]=\left|\det \left[ \mathbf{\nabla m}  (\mathbf{x}) \right] \right|^{-1} \delta_{\mathbf{m}^{-1}(\mathbf{y})}\left( \mathbf{x}\right)                                                                                                             
\end{equation} 
where the substitution of $\mathbf{m}^{-1}(\mathbf{y})$ by $\mathbf{x}$ in the determinant does not affect the distribution.   

To explain the physical meaning of the Jacobian determinants, let consider a small hyperparallelepiped represented by one corner $\mathbf{x}_0$ and a matrix $\delta \mathbf{V}_{0}$ wherein the $j$-th column vector, denoted by $\delta \mathbf{x}^j_0$ is the $j$-th edge arising from $\mathbf{x}_0$. By successive applications of the map, the $j$-th vector evolves according to $\delta\mathbf{x}^j_{n+1}=\mathbf{m}(\mathbf{x}_{n}+\delta \mathbf{x}^j_{n})-\mathbf{m}(\mathbf{x}_{n})$ where $\delta \mathbf{x}^j_n$ and $\delta \mathbf{x}^j_{n+1}$ are the $j$-th edge vector at $\mathbf{x}_n$ and $\mathbf{x}_{n+1} = \mathbf{m}(\mathbf{x}_n)$. 
Expanding the map $\mathbf{m}$ at $\mathbf{x}_{n}$ yields~\cite{ott2002chaos}
\begin{equation}
\mathbf{m}(\mathbf{x}_{n}+\delta\mathbf{x}^j_{n})
=\mathbf{m}(\mathbf{x}_{n})+(\mathbf{\nabla m})^T(\mathbf{x}_{n}) \delta \mathbf{x}^j_{n}+O(\| \delta \mathbf{x}^j_{n}\|^{2})\label{eq:variation_evolution}.\end{equation}
Thus the $j$-th column vector evolves according to
$\delta\mathbf{x}^j_{n+1}=\mathbf{\nabla m}^T(\mathbf{x}_{n})
\delta\mathbf{x}^j_{n}$ at first order. The matrix representing the small
hyperparallelepiped similarly evolves according to
$\delta\mathbf{V}_{n+1}=\mathbf{\nabla m}^T(\mathbf{x}_{n})
\delta\mathbf{V}_{n}$. 
Now, if matrix components at step $0$ are chosen such that $\det(\delta \mathbf{V}_{0})>0$, then this quantity characterizes the initial volume of the small hyperparallelepiped. This volume then evolves according to 
\begin{equation}
 \mathrm{det}(\delta \mathbf{V}_{\ell}) = \det(\delta \mathbf{V}_{0}) \textstyle \prod_{n=0}^{\ell-1} \det \left[ \mathbf{\nabla m} (\mathbf{x}_n) \right]. 
\end{equation}
Thus, the Jacobian determinant product represents the factor by which phase space is compressed or expanded on the $\ell$ successive applications of the map $\mathbf{m}$ starting at $\mathbf{x}_0$. 

A trajectory is defined as an indexed sequence of $L+1$ states in phase space, denoted by $ z =\left\{ \mathbf{x}_0,...,\mathbf{x}_{{L}}\right\}$, \emph{independently from the map} $\mathbf{m}$.  
From~\eqref{eq:jacobian} and~\eqref{eq:jacobian2}, the conditional probability $\mathrm{P}_\mathrm{cond}( z|\mathbf{x}_\ell)$ to generate $z$ by successively applying the map starting from state $\mathbf{x}_\ell$ ($0 \leq \ell \leq L$) is the product of the forward and backward transition probabilities from the corresponding index $\ell$ 
\begin{eqnarray}\label{eq:conditional}
\mathrm{P}_\mathrm{cond}( z|\mathbf{x}_{\ell}) & = & \mathrm{P}_\mathrm{fwd}( z
| \ell , \mathbf{x}_\ell ) \mathrm{P}_\mathrm{bwd}( z | \ell, \mathbf{x}_{\ell})
; \\
 \mathrm{P}_\mathrm{fwd}( z | \ell, \mathbf{x}_{\ell}) &=& \textstyle \prod_{n=\ell}^{L-1} \left|\det \left[ \mathbf{\nabla m}  (\mathbf{x}_n) \right] \right|  \delta_{\mathbf{m}(\mathbf{x}_{n})}\left( \mathbf{x}_{n+1} \right), \nonumber \\
\mathrm{P}_\mathrm{bwd}( z| \ell, \mathbf{x}_{\ell}) & =& \textstyle \prod_{n=0}^{\ell-1} \left|\det \left[ \mathbf{\nabla m}  (\mathbf{x}_n) \right] \right|^{-1} \delta_{\mathbf{m}^{-1}(\mathbf{x}_{n+1})}\left(\mathbf{x}_n\right).  \nonumber 
\end{eqnarray}

For Dirac measures, we have equivalence between $\delta_\mathbf{a}(d\mathbf{b}) d \mathbf{a}$ and $\delta_\mathbf{b}(d\mathbf{a})d\mathbf{b} $. Hence, the conditional probability~\eqref{eq:conditional} can be cast in the general form 
\begin{equation}\label{eq:pcond}
\mathrm{P}_\mathrm{cond}( z|\mathbf{x}_{\ell}) = \mathrm{P}_\mathrm{cond}( z|\mathbf{x}_0) \textstyle \prod_{n=0}^{\ell-1} \det \left[ \mathbf{G} (\mathbf{x}_n) \right]^{-1}, 
\end{equation}
where $\mathbf{G} = (\mathbf{\nabla m })^T \mathbf{\nabla m}$ denotes the Gram matrix. This identity is a key ingredient of the path-sampling schemes described in Sec~\ref{sec:Transition-path-sampling}. We now focus on the map that will be used. 

\section{Verlet map \label{sec:Verlet}}

We consider dynamics governed by an Hamiltonian
$\mathcal{H}(\mathbf{x})=\mathcal{K}(\mathbf{p})+\mathcal{V}(\mathbf{q})$ where
$\mathcal{V}(\mathbf{q})$ and $\mathcal{K}(\mathbf{p})=\frac{1}{2}
\sum_{i=1}^{3N} p_{i}^2/m_i $ denote 
the potential and kinetic energies, respectively. The Hamiltonian gradient $\nabla \mathcal{H} ( \mathbf{x} )$ is written as a column vector $(\nabla_1 \mathcal{H}, ..., \nabla_{6N}\mathcal{H})^T$. 
Hamilton's equation expressed in matricial form reads~\cite{lichtenberg1992regular} 
\begin{equation}
\mathbf{\dot{x}} = \mathbf{J} \nabla \mathcal{H} (\mathbf{x})
\label{eq:hamilton}
\end{equation}
where $\mathbf{J}$ is the canonical skewed-symmetric matrix
\begin{equation}
\mathbf{J}=\left(\begin{array}{cc}
\mathbf{0} & \mathbf{I}_{3N}\\
-\mathbf{I}_{3N} & \mathbf{0}\end{array}\right)\nonumber. 
\end{equation}
and $\mathbf{I}_{3N}$ is the $3N \times 3N$ identity matrix. 
Hamilton's evolution equation is discretized through a map $\mathbf{m}_\tau$ by approximating $\mathbf{x}(n \tau)$ 
with $\mathbf{x}_{n}=\mathbf{m}_\tau(\mathbf{x}_{n-1})$ where $\tau$ is the time-step. Trajectories are $ z=\left\{  \mathbf{x}_0,...,\mathbf{x}_L \right\} \equiv \left\{ \mathbf{x}(0),...,\mathbf{x}(\mathcal{T}) \right\} $ where the total length $\mathcal{T} = {L} \tau $. Let define the vectorial functions $\mathbf{\tilde{q}}$ and $\mathbf{\tilde{p}}$  by $\mathbf{\tilde{q}}(\mathbf{x})=\mathbf{q}$ and $\mathbf{\tilde{p}}(\mathbf{x})=\mathbf{p}$ and let $\circ$ denote the function composition. We write $\widetilde{\mathcal{K}}= \mathcal{K} \circ \mathbf{\tilde{p}} $ and $\widetilde{\mathcal{V}}= \mathcal{V} \circ \mathbf{\tilde{q}} $, making it possible to apply the generalized gradient $\nabla$ on the kinetic and potential energies directly. 
The position-Verlet map is obtained by discretizing Hamilton's equation~\eqref{eq:hamilton} as follows   
\begin{equation} \label{eq:dm_verlet}
\mathbf{m}_\tau=  \left(\mathbf{i} + \tfrac{\tau}{2} \mathbf{J} \nabla \widetilde{\mathcal{K}} \right) \circ \left(\mathbf{i} + {\tau} \mathbf{J} \nabla \widetilde{\mathcal{V}} \right) \circ\left(\mathbf{i} + \tfrac{\tau}{2} \mathbf{J} \nabla \widetilde{\mathcal{K}} \right).  
\end{equation}
where $\mathbf{i}$ is the functional identity : $\mathbf{i}(\mathbf{x})=\mathbf{x}$. 
Defining $\mathbf{q}_{n+1/2}= \left[ \mathbf{\tilde{q}} \circ ( \mathbf{i} + \tfrac{\tau}{2} \mathbf{J} \nabla \widetilde{\mathcal{K}}) \right](\mathbf{x}_{n})$ enables one to decompose $\mathbf{x}_{n+1}= \mathbf{m}_\tau(\mathbf{x}_{n})$ according to the intermediate updates of the splitting~\eqref{eq:dm_verlet}  
\begin{subequations} \label{eq:splitting}
  \begin{align}
 && \mathbf{q}_{n+1/2}   & =  & \mathbf{q}_{n}     & & + & \tfrac{\tau}{2} \nabla_\mathbf{p} \mathcal{K} \left( \mathbf{p}_{n} \right),&& \\ 
 && \mathbf{p}_{n+1}     & =  & \mathbf{p}_n       & & - & \tau \nabla_\mathbf{q} {\mathcal{V}} (\mathbf{q}_{n+1/2}), && \label{eq:q1/2} \\   
 &&\mathbf{q}_{n+1}     & =  & \mathbf{q}_{n+1/2} & & + & \tfrac{\tau}{2} \nabla_\mathbf{p} \mathcal{K} \left(\mathbf{p}_{n+1} \right),&& 
  \end{align}
\end{subequations}
where ${q}_{j,n+1}$ simplifies into ${q}_{j,n} + \tfrac{\tau}{2}({p}_{j,n}+ {p}_{j,n+1})/m_j$. 

\section{Eigenvalue decomposition \label{sec:Eigenvalue-decomposition}}

Verlet maps have three interesting properties~\cite{LRS2010} : (i) time symmetry
(i.e. $\mathbf{m}_{-\tau}=\mathbf{m}_{\tau}^{-1}$);  (ii) reversibility [i.e.
$\mathbf{m}_{-\tau}=\mathbf{r} \circ \mathbf{m}_{\tau}\circ \mathbf{r}$ where
$\mathbf{r}$ is the momentum reversal application:
$(\mathbf{q}^T,\mathbf{p}^T)^T \rightarrow (\mathbf{q}^T,-\mathbf{p}^T)^T$] and
(iii) symplecticity [i.e $(\mathbf{\nabla m_\tau })^T \mathbf{J} \mathbf{\nabla
m_\tau} =  \mathbf{J}$]. 
To prove property (iii), we formally expand the Jacobian matrix for the
position-Verlet map
\begin{eqnarray}
(\mathbf{\nabla m}_\tau)^T & = &  (\mathbf{I} + \tfrac{\tau}{2} \mathbf{J} \widetilde{\mathbf{K}})(\mathbf{I} + \tau \mathbf{J} \widetilde{\mathbf{V}})(\mathbf{I} + \tfrac{\tau}{2}  \mathbf{J} \widetilde{\mathbf{K}}), \label{eq:jacobian_matrix}
\end{eqnarray}
where $\widetilde{\mathbf{K}}$ and $\widetilde{\mathbf{V}}$ are the Hessian matrices of $\widetilde{\mathcal{K}}$, $\widetilde{\mathcal{V}}$, and $\mathbf{I}= \nabla \mathbf{i}^T$. The components are $(\widetilde{\mathbf{K}})_{ij} = \nabla_i \nabla_j \widetilde{\mathcal{K}}$, 
$(\widetilde{\mathbf{V}})_{ij} = \nabla_i \nabla_j \widetilde{\mathcal{V}}$ and $(\mathbf{I})_{ij}= \delta_{ij}$. 
In~\eqref{eq:jacobian_matrix} the Jacobian matrix $(\mathbf{\nabla m}_\tau)^T$ is factored into a product of three symplectic matrices, hence it is also symplectic. Note that the Hessian of the potentiel energy is to be evaluated at $\mathbf{q}_{n+1/2}$ for $(\mathbf{\nabla m}_\tau)^T(\mathbf{x}_n)$ or $(\mathbf{\nabla m}_{-\tau})^T (\mathbf{x}_{n+1})$ as the potential gradient in~\eqref{eq:splitting}.  

An important implication of the symplecticity property is that eigenvalues of the Jacobian matrix occur in reciprocal pairs: if $\mu$ is an eigenvalue, then so is $\mu^{-1}$ with the same algebraic multiplicity. 
To facilitate the eigenvalue decomposition in reciprocal pairs, we perform a change of basis in order to partition the Jacobian matrix into four \emph{commuting} and \emph{symmetric} block matrices. The associated transformation matrix is defined by  
\begin{equation} \nonumber
\mathbf{\Gamma} =  \left(\begin{array}{cc}
\mathbf{K}^{\frac{1}{2}} & \mathbf{0} \\
\mathbf{0} & \mathbf{K}^{-\frac{1}{2}} \end{array}\right). 
\end{equation}
The transformation amounts to switching to mass-weighted coordinates. 
The two block matrices are diagonal and have dimension $3N \times 3N$ with entries 
${(\mathbf{K}^{\alpha})}_{ij} = m_i^{-\alpha} \delta_{ij} $. Note that $\mathbf{K}^{1}$ is the Hessian matrix of the kinetic energy $\mathcal K$. 
After expressing the various matrices composing the Jacobian matrix (Eq.~\eqref{eq:jacobian_matrix}) in a basis with mass-weighted coordinates, the latter one becomes $\mathbf{\Gamma}^{-1} (\mathbf{\nabla m }_\tau)^T \mathbf{\Gamma}$ and writes explicitly
\begin{equation}  \label{eq:dm_hamilt}
\left(\begin{array}{cc}
\mathbf{C} & \tfrac{\tau}{2} (\mathbf{C}+\mathbf{I}_{3N}) \\
 \tfrac{2}{\tau} (\mathbf{C}-\mathbf{I}_{3N})& \mathbf{C} \end{array}\right), 
\end{equation}
with $\mathbf{C}=\mathbf{I}_{3N}-\frac{\tau^2}{2} \mathbf{K}^\frac{1}{2}\mathbf{V}\mathbf{K}^\frac{1}{2}$. The mass-weighted Hessian matrix of the potential energy $\mathcal{V}$ is symmetric since we have
\begin{equation}
\nonumber
(\mathbf{K}^\frac{1}{2}\mathbf{V}\mathbf{K}^\frac{1}{2})_{ij} = \frac{1}{\sqrt{m_i m_j}} \frac{\partial^{2}V}{\partial{q}_i\partial {q}_j}= (\mathbf{K}^\frac{1}{2}\mathbf{V}\mathbf{K}^\frac{1}{2})_{ji}. 
\end{equation}
Hence, $\mathbf{K}^\frac{1}{2}\mathbf{V}\mathbf{K}^\frac{1}{2}$ admits real eigenvalues, denoted by $\omega_j^2$ and sorted here in ascending order. An unstable normal mode is characterized by a negative eigenvalue $\omega_j^2$ in which case the eigenfrequency $\omega_j$ is pure imaginary. A stable normal mode is characterized by a positive eigenvalue $\omega_j^2$ in which case the eigenfrequency $\omega_j$ is real. The eigenvalues of $\mathbf{C}$ are $c_j=1-\omega_j^2\tau^2/2$. 

The commutative subring formed by the block matrices $\mathbf{C}$, $\frac{\tau}{2}(\mathbf{C} + \mathbf{I}_{3N})$ and $\frac{2}{\tau}(\mathbf{C} - \mathbf{I}_{3N})$ within the partitioned Jacobian matrix~\eqref{eq:dm_hamilt} allows us to factorize its characteristic polynomial (secular equation) as follows ($\mu \in \mathbb{C}$)
\begin{equation}\label{eq:det_lambda}
\det \left[ \left(\mu \mathbf{I}_{3N} - \mathbf{C} \right)^2 +  \mathbf{I}_{3N} - \mathbf{C}^2  \right]=0.
\end{equation}
Diagonalizing the matrices of~\eqref{eq:det_lambda} in the real eigenbasis of $\mathbf{C}$ enables one to decompose the secular equation into the product of the following second order equations 
\begin{equation}
 (\mu - c_j)^2 + 1- c_j^2 =0. \label{eq:secondorder}
\end{equation}
If $\mu_j$ is a root of~\eqref{eq:secondorder}, then Vieta's formula implies that $\mu_j^{-1}$ is the second root and we have ($i = \sqrt{-1}$) 
\begin{equation} \nonumber
 \mu_j^{\pm 1}= c_j \pm i \sqrt{1-c_j^2 }. 
\end{equation}
For convenience, we introduce an effective frequency $\widetilde{\omega}_j$ to express the matching pairs of roots as 
\begin{equation} \nonumber
 \mu_j^{\pm 1}= \exp \left( \pm i \widetilde{\omega}_j \tau \right). 
\end{equation}

For $c_j > 1$ (i.e. $\omega^2_j < 0$), $\omega_j$ and $\widetilde{\omega}_j$ are both pure imaginary numbers. 
We have $\cosh (i \widetilde{\omega}_j \tau) = c_j $ and $\sinh(i\widetilde{\omega}_j \tau) = - \sqrt{|1-c_j^2|} < 0$. Consequently, $i \widetilde{\omega}_j < 0$ and $\sinh(i\widetilde{\omega}_j \tau /2) = - \sqrt{\tfrac{1}{2} [\cosh(i\widetilde{\omega}_j \tau)-1] } = - \sqrt{ \tfrac{1}{2} (c_j -1)}$. Substituting $1-\omega_j^2 \tau^2 /2 $ for  $c_j$ in the last square root and setting ${\omega}_j$ to the value $i | \omega_j|$ yields $\sinh (i \widetilde{\omega}_j \tau/{2} )= i \omega_j \tau/2 $.  
The corresponding Jacobian eigenvalues $\mu_j$ and $\mu_j^{-1}$ are real. The projections of $\delta\mathbf{x}$ along the eigendirection of $\mu_j$ and $\mu_j^{-1}$ are contracting and expanding, respectively. The occurrence of an unstable normal mode characterizes configurations in saddle regions. 
For $c_j =1$, we have $\omega_j= \widetilde{\omega}_j=0$. 
In any energy surface exhibiting periodic boundary conditions (as will be the case in
Sec.~\ref{Numerical results}) there is a zero frequency for each normal mode associated 
with a translational symmetry.  
For $-1 < c_j < 1$, (i.e. $0 < \omega_j^2 < 4/\tau^2$), $\omega_j$ and
$\widetilde{\omega}_j$ are both real positive. 
We have $\cos (\widetilde{\omega}_j \tau) = c_j $ and $\sin (\widetilde{\omega}_j{\tau})= \sqrt{1-c_j^2} >0 $. Since $\omega_j$ is chosen positive, we have $\sin (\widetilde{\omega}_j{\tau}/{2} )= \omega_j {\tau}/{2}$. 
In this case, $(\mu_{j},\mu_{j}^{-1})$ is a conjugate pair in the unitary circle of the complex plane (with unit modulus). 
This favors oscillations of the projection of $\delta\mathbf{x}$ into the plane containing the two corresponding eigendirections. 
States in metastable basins have all their normal modes stable ($\omega_j^2$ positive). 
For $c_j \leq -1$ (i.e. $\omega^2_j \geq 4\tau^{-2}$), $\widetilde{\omega}_j$ is pure imaginary 
while the corresponding normal mode is stable (${\omega}_j$ is real positive). 
The {\em effective } mode of the discretized dynamics is unstable but becomes stable again if the time-step $\tau$ is decreased below $2/\omega_j$. The situation $c_j \leq -1$ thus involves numerical instability and will not be covered in the following as admissible time-steps in molecular dynamics should be a fraction of the fastest vibration period (see the linear stability analysis in Ref.~\onlinecite{LRS2010}). 

Hence, assuming $\omega^2_j \in \mathbb{R}$ and $\omega^2_j < 4\tau^{-2}$, we may choose 
\begin{equation}
\widetilde{\omega}_j = \frac{2}{\tau} \left[ \arcsin \circ \mathrm{Re} \left(\omega_j \frac{\tau}{2} \right) + i \times \mathrm{arsinh}\circ \mathrm{Im} \left(\omega_j \frac{\tau}{2}\right) \right] \label{eq:omega_tilde}, 
\end{equation}
where $\mathrm{Re}$ and $\mathrm{Im}$ denote the real and imaginary parts of a complex number and 
$\circ$ still denotes the functional composition. 
Equation~\eqref{eq:mu_reduite} generalizes in the complex plane as follows 
\begin{eqnarray}
\label{eq:mu_reduite}
\widetilde{\omega}_j &=& 2 (i \tau )^{-1} \ln \left( \sqrt{1- \omega_j^2 \tau  ^2/4 } + i \omega_j \tau /2  \right), 
\end{eqnarray}
provided the principal branch of the multi-form complex function is delimited by the cutting segments $]-\infty, -2\tau^{-1} [$ and $] 2\tau^{-1}, +\infty [$. 
The linear approximation of the effective frequency $\widetilde{\omega}_{j}$ is
$\omega_{j}$. To detail the higher terms, let first
derivate~\eqref{eq:mu_reduite} with respect to $\omega_j$ and then expand the
result in Taylor series ($\omega^2_j \in \mathbb{R}$ and $\omega^2_j <
4\tau^{-2}$):
\begin{equation}
 \frac{\mathrm{d} \widetilde{\omega}_j}{\mathrm{d} {\omega}_j} = 1  \left/ \sqrt{1-\omega_j^2 \tau^2 /4} \right.
= \sum_{k=0}^{\infty} \tbinom{2k}{k}  \left(\tfrac{1}{4} \omega_j \tau \right)^{2k} \label{eq:derivative} .
\end{equation}
The effective frequency $\widetilde{\omega}_j$ is obtained by integrating the function $\omega \rightarrow \sum_{k=0}^{\infty} \tbinom{2k}{k}  \left(\tfrac{1}{4} \omega \tau \right)^{2k}$ from 0 to $\omega_j$ 
\begin{equation} \label{eq:series}
 \widetilde{\omega}_j = \omega_j \sum_{k=0}^{\infty} \tfrac{1}{2 k+1} \tbinom{2k}{k} \left( \tfrac{1}{4} \omega_j \tau \right)^{2k}
\end{equation}
where we used the fact that $\widetilde{\omega}_j$ and ${\omega}_j$ both take value zero at $c_j=1$. 
The series is real positive whatever $\omega_j^2 \in \mathbb{R}$ and $\omega_j^2 < 4\tau^{-2}$. 
The non-linear contributions absorbed into $\widetilde{\omega}_j$ contain even powers of $\tau$ exclusively, a property reflecting the oddity of the circular and hyperbolic inverse sines in~\eqref{eq:omega_tilde}. 
The first terms of the series are
\begin{eqnarray} \nonumber
\widetilde{\omega}_j & =  {\omega}_j & \left[  1 + \frac{\tau^2 }{24} \omega_j^2 +\frac{3\tau^4 }{640}\omega_j^4  + \frac{5\tau^6}{7168} \omega_j^6 + \cdots \right].  \label{eq:puiseux}
\end{eqnarray}
In practice, an appropriate truncation of~\eqref{eq:series} is used to avoid numerical round-off errors when evaluating inverse sine functions close to 0. 

Let us now characterize the extremal moduli of the eigenspectrum. The imaginary parts $\mathrm {Im} \left({\omega}_{j} \right)$ form a descending series,  
as for the imaginary parts $\mathrm {Im} \left( \widetilde{\omega}_{j} \right)$
since $\mathrm{d}\widetilde{\omega}_j/ \mathrm{d}\omega_j$ is real
positive in~\eqref{eq:derivative}. Conversely, the series formed by the moduli $|\mu_{j}|= \exp \circ
\mathrm{Im} (-\widetilde{\omega}_{j} \tau)$ is ascending for the exponential
function being increasing on $\mathbb{R}$ with $\mathrm{Im}
(-\widetilde{\omega}_{j})$.  Therefore, the smallest and largest moduli are
$|\mu_1|$ and $|\mu_1|^{-1}$, two quantities always associated with the lowest
eigenvalue $\omega^2_1$. Physically, the matching pairs of Jacobian eigenvalues
($\mu_{j},\mu_{j}^{-1}$) are local Lyapunov numbers. The local Lyapunov
exponent, defined by 
\begin{equation} \nonumber
-\tau^{-1} \ln | \mu_1  | = \mathrm{Im} (\widetilde{\omega}_1) = \frac{2}{\tau}  \mathrm{arsinh} \circ \mathrm{Im} (\frac{\tau}{2}{\omega}_{1})
\end{equation}
is associated to the maximum expansion rate of a perturbation $\mathbf{\delta x}_n$ at current state $\mathbf{x}_n$. 
Lyapunov exponents rather correspond to the expansion rate in the infinite time limit $\tau \rightarrow + \infty$.~\cite{ott2002chaos} 

Lyapunov exponents measure the sensitivity of dynamical systems to small changes
in initial conditions and, for this reason, are widely employed for
characterizing the dynamical properties of many nonlinear and Hamiltonian
systems.~\cite{lichtenberg1992regular} For instance, by monitoring the Jacobian
eigenvalues of the velocity-Verlet map and averaging the logarithm of the
eigenvalue moduli exceeding one,~\cite{hinde1992chaos,amitrano1992probability,
hinde1993chaotic,calvo1998chaos,W2003} the chaotic or regular nature of
dynamical trajectories could be identified and related to the occurrence or
absence of structural transitions in small Lennard-Jones clusters. 
In two recent investigations, the Jacobian eigenvalue spectrum itself 
has been successfully used to enhance the occurrence of rare chaotic and regular 
trajectories in computer simulations of dynamical systems.~\cite{TK2006,geiger2010identifying}
In the Lyapunov-weighted dynamics (LWD) developed by Tailleur and 
Kurchan~\cite{TK2006} (see also Refs.~\onlinecite{TTK2006} and~{picciani2011simulating}), a set of small 
vectors evolves according to the Jacobian 
equation~\eqref{eq:variation_evolution} coupled to a birth-death process 
consisting in multiplicating or deleting vectors depending on their norms. In 
the Lyapunov-weighted path-sampling (LWPS) method developed by Geiger and 
Dellago,~\cite{geiger2010identifying} the  birth-death process is replaced by a 
Metropolis algorithm sampling short dynamical trajectories. The Metropolis 
acceptance rate of LWPS includes a biasing factor proportional to the relative 
Lyapunov indicator~\cite{sandor2004relative} of the involved trajectories.
The ability of LWPS to identify rare trajectories suggests that this approach is
well suited to the estimation of time-correlation functions associated with
thermally activated processes.

\section{Sampling dynamically activated events \label{sec:Transition-path-sampling}}

We implement a variant form of LWPS. Instead of resorting to the relative Lyapunov indicator we use $\textstyle \prod_{n=0}^{{L}-1} | \mu_{1,n+1/2} |$, the product of the smallest eigenvalue moduli along trajectory $z$, to bias the importance function. The lowest eigenvalue $\mu_{1,n+1/2}$ refers to Jacobian matrices $(\nabla \mathbf{m}_\tau)^T(\mathbf{x}_n)$ or $(\nabla \mathbf{m}_{-\tau})^T(\mathbf{x}_{n+1})$. The logarithm of the biasing quantity reads 
\begin{eqnarray} \nonumber
 \mathcal{L}( z) = -2 \sum_{n=0}^{{L}-1} \mathrm{arsinh} \circ  \mathrm{Im}\left( {\omega}_{1}(\mathbf{q}_{n+1/2}) \frac{\tau}{2} \right) \le 0. 
\end{eqnarray} 
$\mathcal{L}(z)$ is hereafter called ``activation indicator''. By definition, a trajectory is said to be inactive when its indicator is zero and to be active when it is strictly negative, a negatively curved portion of the energy surface being visited in the latter case. At variance, the relative Lyapunov indicator does not always discriminate active trajectories from inactive ones. 

The lowest eigenvalue $\omega^2_1 \left( \mathbf{q}_{n+1/2} \right)$ is calculated using the 
Lanczos algorithm \cite{lanczos1961applied} as in activation-relaxation 
techniques.~\cite{PhysRevLett.77.4358,MB1998,cances2009,machado2011} At each implementation, the algorithm finds the 
lowest (or highest) eigenvalue and eigenvector of any matrix at a reduced computational cost, 
by constructing a Krylov basis of size $d$ and diagonalizing the associated $d \times d$ matrix where $d$ is small 
(see for example appendix A of Ref.~\cite{marinica2011energy} for details). 
As pointed out in Ref.~\onlinecite{cosminart}, using a Krylov basis of size $d=15$ guarantees 
the convergence of eigenvalue $\omega^2_1 \left( \mathbf{q}_{n+1/2} \right)$. 
Herein, we have decreased the basis size to $d=8$ to increase
numerical efficiency. At 
each implementation, we verify that the Lanczos solution is stable; if it is not the case, 
the calculation is repeated until the solution is converged up to $10^{-3}\,eV / \AA ^2$. 
The lowest eigenvalue $\omega_{1}$ is on average obtained after 1.5 iterations
or equivalently 24 force evaluations. Compared to calculations based on the relative Lyapunov indicator, the present
approach ensures stability and high accuracy in the evaluation of $\omega_{1}$
which is essential in the estimation of the activation indicator. 

The Monte Carlo method consists in sampling the trajectorial phase space endowed with parametrized probability density
\begin{equation} 
\mathrm{P}^{\theta}_{A}( z)=\mathrm{P}_\mathrm{cond}( z|\mathbf{x}_0) h_{A}(\mathbf{x}_{0})
\exp\left[f_\theta- u_\theta( z) \right]  \label{eq:9}
\end{equation}
where $f_{\theta}$, the reduced free energy of the biased trajectory ensemble, acts as a normalizing constant. 
The $\theta$ parameter controls the biasing strength of the activation indicator in the path action 
\begin{equation}
 \label{eq:bias}
 u_\theta( z) = \beta \mathcal{H}(\mathbf{x}_0) + \theta \mathcal{L}( z). 
\end{equation}
Note that $\mathrm{P}^{\theta}_{A}( z)=0$ if $\mathbf{x}_0 \notin A$ : paths are all initiated from the reactant basin whatever  $\theta$. Besides, $f_\theta$ in~\eqref{eq:9} depends not only on $\theta$ but also on $A$ and the inverse temperature $\beta$. 
In the biased path ensemble, increasing the value of $\theta$ decreases $\langle \mathcal{L}( z) \rangle_\theta$ 
the ensemble average of the activation indicator. It thus enhances the
occurrence of excursions 
into unstable regions of the potential energy landscape and the occurrence of
active trajectories within the basin and the reactive trajectories crossing a
saddle region. Conversely, decreasing the value $\theta$ favors the occurrence
of inactive trajectories. 

The present ensemble is similar to the $s$-ensemble used by Chandler and co-workers~\cite{elmatad:2010} to bias the occurence of dynamical heterogeneities in glasses. Their biasing field $s$ couples to a dynamical activity defined to be an excitation indicator (number of spin flips in a trajectory) in the same way as the conjugate parameter $\theta$ couples to $\mathcal{L}$. 

The distribution $\mathrm{P}^{\theta}_{A}$ is approximated by a Markov chain of \emph{M} steps constructed using a Metropolis Monte Carlo method. As in TPS or Ref.~\onlinecite{elmatad:2010}, the scheme consists in repeatedly attempting shooting and shifting moves. 

\subsection{Shooting moves}

A shooting move consists in performing the following operations: 
\begin{enumerate}[(i)]
\item select a state $\mathbf{x}_{\ell}= 
\left(\mathbf{q}_\ell^T,{\mathbf{p}}_\ell^T\right)^T$ in the current trajectory
by drawing an integer $\ell$ uniformly in $[0, L]$;

\item generate trial momenta $\tilde{\mathbf{p}}_\ell$ from current momenta
$\mathbf{p}_\ell$ and define trial state $\tilde{\mathbf{x}}_\ell =
\left(\mathbf{q}_\ell^T,\tilde{\mathbf{p}}_\ell^T\right)^T$; 

\item construct the trial trajectory $\widetilde{z}$ by propagating two trajectory segments from $\widetilde{\mathbf{x}}_\ell$, one backward of duration $\ell \tau$ and the other one forward of duration $\mathcal{T}-\ell\tau$; 

\item compute the acceptance probability of $\tilde{z}$
\begin{equation} 
\label{eq:Pacc_shoot}
\hspace{0.75cm} \mathrm{P}_{\mathrm{acc}}\left[\widetilde{ z}\leftarrow { z}\right]=\min \left\{ 1,h_{A}(\widetilde{\mathbf{x}}_{0}) \exp \left[u_\theta( z) - u_\theta (\widetilde{ z})\right] \right\};
\end{equation}

\item draw a random number $\mathsf{Rand1} \in (0,1]$ and accept the trial trajectory $\widetilde{ z}$ if $\mathsf{Rand1} \leq \mathrm{P}_{\mathrm{acc}}\left[\widetilde{ z}\leftarrow z\right] $ otherwise the trial trajectory is rejected and the current trajectory is duplicated in the Markov chain.
\end{enumerate}
As detailed below, acceptance probability~\eqref{eq:Pacc_shoot} formally corresponds to the traditional Metropolis acceptance rate
\begin{equation}\label{eq:formal}
\mathrm{P}_{\mathrm{acc}}\left[\widetilde{ z}\leftarrow z\right]=\min\left\{ 1,\frac{\mathrm{P}_\mathrm{gen}\left[{ z}\leftarrow \widetilde{ z}\right]\mathrm{P}^{\theta}_A \left[\widetilde{ z}\right]}{\mathrm{P}_\mathrm{gen}\left[\widetilde{ z}\leftarrow{ z}\right]\mathrm{P}^{\theta}_A \left[ z\right]} \right\}. 
\end{equation}
where $\mathrm{P}_\mathrm{gen} \left[ \widetilde{z} \leftarrow {z}\right]$ is the probability to generate $\tilde{z}$ from $z$ and vice versa for $\mathrm{P}_\mathrm{gen}\left[{z}\leftarrow \widetilde{z}\right]$. Consequently, the probability fluxes $\mathrm{P}_{\mathrm{acc}}\left[\widetilde{ z}\leftarrow {z}\right] 
\mathrm{P}_\mathrm{gen}\left[\widetilde{ z}\leftarrow { z}\right]
\mathrm{P}^{\theta}_A\left[ z\right]$ from $z$ to $\tilde{z}$
and \\
$\mathrm{P}_{\mathrm{acc}}\left[{ z}\leftarrow \widetilde{z}\right] 
\mathrm{P}_\mathrm{gen}\left[{ z}\leftarrow \widetilde{ z}\right]
\mathrm{P}^{\theta}_A\left[ \widetilde{z}\right]$ from $\tilde{z}$ to ${z}$
are balanced, and, shooting moves sample the equilibrium distribution
$\mathrm{P}^\theta_{A}$. 

To show that the formal Metropolis rate~\eqref{eq:formal} and the rate~\eqref{eq:Pacc_shoot} used in practice are equivalent, let us detail the probability of generating the trial trajectory $\widetilde{ z}$ from $ z$ 
\begin{equation}\label{distrp}
\mathrm{P}_\mathrm{gen}[\widetilde{ z} \leftarrow { z}]= \mathrm{P_{cond} }(\widetilde{ z}|\widetilde{\mathbf{x}}_{\ell}) p\left(\widetilde{\mathbf{p}}_\ell\leftarrow{\mathbf{p}}_\ell\right) \frac{1}{L+1}. 
\end{equation} 
The quantities $(L+1)^{-1}$ and $p\left(\widetilde{\mathbf{p}}_\ell\leftarrow{\mathbf{p}}_\ell\right)$ are the probabilities to draw $\ell$ in step (i) and to generate the trial momenta $\widetilde{\mathbf{p}}$ from $\mathbf{p}$ in step (ii) using a simple procedure proposed by Stoltz~\cite{stoltz2007path} based on an Ornstein-Uhlenbeck process on the momenta
\begin{equation} \label{eq:stoltz}
\widetilde{\mathbf{p}}_\ell=\epsilon\mathbf{p}_\ell+\sqrt{1-\epsilon^{2}}\delta\mathbf{p}_\ell. 
\end{equation}
Here $\epsilon$ is a mixing parameter and each $\delta {p}_{i,\ell}$ variate is drawn from a Gaussian
distribution of variance $m_i /\beta$. 
Thus, the probability of generating $\widetilde{\mathbf{p}}_\ell$ from $\mathbf{p}_\ell$ reads
\begin{equation}\nonumber
p\left(\widetilde{\mathbf{p}}_\ell \leftarrow {\mathbf{p}}_\ell \right)= K_\epsilon \exp\left[-\beta\sum_{i=1}^{3N}\frac{(\widetilde{p}_{i,\ell}-\epsilon p_{i,\ell})^2}{2m_i (1-\epsilon^{2}) }\right], 
\end{equation} 
where $K_\epsilon = \prod_{i=1}^{3N}{\sqrt{\beta/[2m_i(1-\epsilon^{2})\pi]}}$. This form ensures that the probability fluxes between the current and perturbed momenta are balanced
\begin{equation}\label{eqht}
\frac{p\left({\mathbf{p}}_\ell\leftarrow \widetilde{\mathbf{p}}_\ell\right)} {p\left(\widetilde{\mathbf{p}}_\ell\leftarrow{\mathbf{p}}_\ell\right)} 
= \frac{\exp\left[-\beta \mathcal{K}(\mathbf{p}_\ell)\right]}
       {\exp\left[-\beta \mathcal{K}(\widetilde{\mathbf{p}}_\ell)\right] }. 
\end{equation} 

As a corollary of the symplecticity property satisfied by $(\nabla
\mathbf{m})^T$, the Jacobian and Gram determinants are equal to one whatever $0
\leq \ell \leq L$: the dynamics preserves the volume of any comoving
infinitesimal hyperparallelepiped. As a result, the current and trial path
probabilities conditioned on $\widetilde{\mathbf{x}}_\ell$ and
$\mathbf{x}_\ell$ respectively, [see equation~\eqref{eq:pcond}],
become 
\begin{subequations}
\label{eq:pcond_simplification}
\begin{align}
\mathrm{P}_\mathrm{cond}( z|\mathbf{x}_{\ell}) = \mathrm{P}_\mathrm{cond}( z|\widetilde{\mathbf{x}}_0), \\
\mathrm{P}_\mathrm{cond}( \widetilde{z}|\widetilde{\mathbf{x}}_{\ell}) = \mathrm{P}_\mathrm{cond}( \widetilde{z}|\widetilde{\mathbf{x}}_0). 
\end{align}
\end{subequations}
Combining~\eqref{distrp},~\eqref{eqht} and~\eqref{eq:pcond_simplification} yields the following ratio 
\begin{equation}\label{eq:ratio}
\frac{\mathrm{P_{gen}}\left({ z}\leftarrow \widetilde{ z}\right) }{\mathrm{P_{gen}}\left(\widetilde{ z} \leftarrow{ z}\right)}= \frac{\mathrm{P_{cond} }({ z}|\mathbf{x}_0) \exp\left[-\beta \mathcal H({\mathbf{x}}_\ell)\right]}{\mathrm{P_{cond} }(\widetilde{ z}|\widetilde{\mathbf{x}}_0)\exp\left[-\beta \mathcal H(\widetilde{\mathbf{x}}_\ell)\right]}
\end{equation} 
where we use the fact that the potential energy of the trial and current states at the shooting index are equal to each other.  
Inserting the ratio~\eqref{eq:ratio} in the Metropolis acceptance probability~\eqref{eq:formal} and resorting to the time symmetry property of the Verlet map yields the explicit form~\eqref{eq:Pacc_shoot} of (iv). 
The small numerical drift on the Hamiltonian resulting from the integration of the position-Verlet scheme from  ${\mathbf{x}}_{0}$ to ${\mathbf{x}}_\ell$ ($\mathcal{H}(\mathbf{x}_\ell) \backsimeq \mathcal H({\mathbf{x}}_0)$) 
and from $\widetilde{\mathbf{x}}_\ell$ to $\widetilde{\mathbf{x}}_0$ ($\mathcal H(\widetilde{\mathbf{x}}_{0}) \backsimeq \mathcal H(\widetilde{\mathbf{x}}_\ell)$) has not been included in the Metropolis acceptance rate. This approximation should not affect the temperature of the initial equilibrium state as each new momenta are generated from the exact Maxwell-Boltzmann distribution (using an Ornstein-Uhlenbeck process). The drift essentially affects the exactitude of Hamiltonian dynamics and must be kept small 
by using small enough a time-step.

\subsection{Shifting moves}\label{waste-recycling}

A shifting move is built upon a multiple proposal procedure~\cite{AM2010} and consists in performing the following operations
\begin{enumerate}[(i)]
\item draw a random integer $\nu$ in interval $[0 ,  L ]$;
\item propagate a first trajectory segment backward from $\mathbf{x}_{0}$ for $\nu$ steps and a second trajectory segment forward from $\mathbf{x}_{L}$ for $L-\nu$ steps; join them to $z$ to form $\mathbf{\zeta}=\{ \widetilde{\mathbf{x}}_{n} \} _{0\leq n\leq2L}$, the joint trajectory where $\widetilde{\mathbf{x}}_{n}$ now denotes shifted state $\mathbf{x}_{n-\nu}$; 

\item for each possible trial trajectory $\tilde{z}_j = \{ \widetilde{\mathbf{x}}_{j+n} \}_{0 \leq n \leq L}$ contained in the joint trajectory $\zeta$, compute the associated selection probability
\begin{equation}\label{pisel}
\mathrm{P}^\theta_\mathrm{sel}(\widetilde{ z}_{j}|\mathbf{\zeta})= h_{A}(\widetilde{\mathbf{x}}_{j})\exp\left[U_\theta(\zeta)-u_{\theta}(\widetilde{ z}_j)\right]
\end{equation}
where the joint action $U_\theta(\zeta)$ is defined by 
\begin{equation}\nonumber
\exp \left[-U_\theta (\zeta ) \right]=\sum_{j=0}^{{L}}h_{A}(\widetilde{\mathbf{x}}_{j})\exp\left[-u_{\theta}(\widetilde{ z}_j)\right].
\end{equation}

\item draw a random number $\mathsf{Rand2}\in (0,1]$ and evaluate the lowest integer $\ell$ such that $\sum_{j=0}^{\ell} \mathrm{P}^\theta_\mathrm{sel}(\widetilde{ z}_{j}|\mathbf{\zeta}) > \mathsf{Rand2}$. 

\item add $\widetilde{ z}_{\ell}$ to the Markov chain.

\end{enumerate}

To show that the shifting moves sample the distribution $\mathrm{P}_A^\theta$, we will prove that the transition probability fluxes between $z=\widetilde{z}_\nu$ and the selected trajectory $\widetilde{z}_\ell$ are balanced. 

The conditional probability to construct joint trajectory $\mathbf{\zeta}$ starting from the current trajectory $\widetilde{ z}_j$ reads
\begin{eqnarray} \nonumber
\mathrm{P}_\mathrm{cond}(\mathbf{\zeta}|\widetilde{ z}_j) & = & \mathrm{P}_\mathrm{bwd}(\mathbf{\zeta}|j,\widetilde{\mathbf{x}}_{j}) \mathrm{P}_\mathrm{fwd} (\zeta |j+L,\widetilde{\mathbf{x}}_{j+L})/(L+1) \\
& = &  {\mathrm{P}_\mathrm{cond}({\zeta}|\widetilde{\mathbf{x}}_{j})} / \left[ \mathrm{P}_\mathrm{cond} ( \widetilde{ z}_j |\widetilde{\mathbf{x}}_{j})(L+1) \right] \nonumber
\end{eqnarray}
where $(L+1)^{-1}$ is the probability to choose a particular $\nu$. 
The probability of each trial path $\widetilde{ z}_j$ is $\mathrm{P}_A^\theta(\widetilde{ z}_j)$ with the characteristic function $h_A$ acting upon the initial state $\widetilde{\mathbf{x}}_{j}$ [see Eq.~\eqref{eq:9}]. 

The marginal probability associated with a joint trajectory $\zeta$ reads 
\begin{equation}\nonumber
\mathrm{P}_{\mathrm{marg}}^{\theta}(\mathbf{\zeta})=\sum_{j=0}^{{L}} \mathrm{P_{cond}}(\mathbf{\zeta} | \widetilde{ z}_{j}) \mathrm{P}^{\theta}_{A}(\widetilde{ z}_{j}) .
\end{equation}
which, for the position-Verlet map, simplifies into 
\begin{equation}\label{pmarg}
\mathrm{P}_{\mathrm{marg}}^{\theta}(\mathbf{\zeta}) = \mathrm{P}_\mathrm{cond}(\mathbf{\zeta}|\widetilde{\mathbf{x}}_{0}) \exp \left[ f_\theta-U_\theta (\zeta) \right]. 
\end{equation}
wherein the indicator function $h_A$ is hidden in the joint action $U_\theta (\zeta)$. 
The probability~\cite{AM2010} of selecting $\widetilde{ z}_{j}$ knowing the ``joint'' trajectory $\mathbf{\zeta}$ is a posterior likelihood given by Bayes relation
\begin{equation}\label{bayes}
\mathrm{P}_\mathrm{sel}^\theta(\widetilde{ z}_{j}|\mathbf{\zeta})=\frac{\mathrm{P}_\mathrm{cond}(\mathbf{\zeta}|\widetilde{ z}_{j})\mathrm{P}^{\theta}_{A}(\widetilde{ z}_{j})}{\mathrm{P}_\mathrm{marg}^{\theta}(\mathbf{\zeta})}. 
\end{equation}
The ratios of conditional probabilities that appear when
developing~\eqref{bayes} simplifies into 1 for the position-Verlet scheme (see~\eqref{eq:pcond_simplification}). As a
result, the posterior probability simplifies to~\eqref{pisel}. 

The multi-proposal shifting move consists of generating the joint trajectory $\mathbf{\zeta}$ from the current path $ z$ using $\mathrm{P}_\mathrm{cond}(\mathbf{\zeta}| z)$ and then selecting the next path of the Markov chain among the $\widetilde{ z}_j$ proposals included in $\zeta$ using posterior probability $\mathrm{P}_\mathrm{sel}^\theta$. These shifting moves leave the probability distribution $\mathrm{P}^{\theta}_{A}$ invariant because they satisfy the detailed balance condition ($ z=\widetilde{ z}_{\nu}$)
\begin{equation}\nonumber
\mathrm{P}_\mathrm{sel}^\theta(\widetilde{ z}_{j}|\mathbf{\zeta})\mathrm{P}_\mathrm{cond}(\mathbf{\zeta}|\widetilde{ z}_{\nu})\mathrm{P}^{\theta}_{A}(\widetilde{ z}_{\nu})=\mathrm{P}_\mathrm{sel}^\theta(\widetilde{ z}_{\nu}|\mathbf{\zeta})\mathrm{P}_\mathrm{cond}(\mathbf{\zeta}|\widetilde{ z}_j)\mathrm{P}^{\theta}_{A}(\widetilde{ z}_{j})
\end{equation}
where 
$\mathrm{P}_\mathrm{sel}(\widetilde{ z}_{\nu}|\mathbf{\zeta})\mathrm{P}_\mathrm{cond}(\mathbf{\zeta}|\widetilde{ z}_j)$ is the probability to transit from $\widetilde{ z}_{j}$ to $ z$, and, conversely, $\mathrm{P}_\mathrm{sel}^\theta(\widetilde{ z}_{j}|\mathbf{\zeta})\mathrm{P}_\mathrm{cond}(\mathbf{\zeta}|\widetilde{ z}_\nu)$ is the probability to transit from $ z$ to $\widetilde{ z}_{j}$. 

A Monte Carlo cycle consists in performing a shooting move followed by a shifting one. 
The constructed sample consists of the trajectories obtained after shifting and is denoted by $\left\{z^1,\cdots ,z^m , \cdots , z^M \right\}$. The associated sample of joint trajectories is denoted by $\left\{z^1,\cdots ,z^m , \cdots , z^M \right\}$. 

\section{Estimating trajectory observables~\label{sec:Reaction-rate-constants-calculation}}

In the following, we employ a second control parameter $\alpha$ which formally may adopt the full range of possible values of $\theta$. 
The intent of the two parameters are distinct however. The parameter $\alpha$ is meant to indicate the target distribution, i.e. the thermodynamic state that we are interested in measuring.  As such the meaningful value of $\alpha$ will strictly be 0 in the context of time-correlation functions. Measurements performed at $\alpha \neq 0$ nevertheless enable one to assess the correctness of the results in term of fluctuations. Unlike the static $\alpha$ parameter, the $\theta$ parameter refers to the various distributions $\mathrm{P}^\theta_A$ that our Monte Carlo algorithm has sampled. 

\subsection{Standard estimator}

The ensemble average of a trajectory observable $\mathcal{O}$ 
\begin{eqnarray}
 \langle \mathcal{O} \rangle_\alpha & = & \int \mathcal{O}( z)  \mathrm{P}_A^\alpha( z) \mathcal{D} z   \label{subeq:prior}
\end{eqnarray}
with respect to distribution $\mathrm{P}_A^\alpha$ given a sample of $M$ trajectories $\{ z^m \}_{1 \leq m \leq M}$ distributed according to $\mathrm{P}_A^\theta$ can be approximated using the following online estimator
\begin{equation}\label{est_corrf_m_unb}
  I^\mathrm{std}_{\alpha,\theta}\left(\mathcal{O}\right)=\sum_{m=1}^{M} 
\frac{\mathcal{O}( z^{m}) \exp\left[\hat{f}_\alpha - u_\alpha( z^{m}) \right] }{M \exp \left[ \hat{f}_\theta - u_\theta( z^{m})  \right] } .
\end{equation}
The reduced free energy estimates $\hat{f}_\alpha$ and  $\hat{f}_\theta$ act as normalizing constants and may be obtained, up to an additive constant, by solving the equation $  I^\mathrm{std}_{\alpha,\theta}\left[ 1 \right]=1$. 

\subsection{Waste-recycling \label{subsec:wr}}

The statistical variance of the standard estimator can be reduced if one recycles information relative to unselected trial moves.~\cite{wr1,wr2} 
Here, we will include information about all the unselected shifting moves contained in the joint trajectory $\zeta$ (see Sec.~\ref{waste-recycling}). The desired waste-recycling estimator of trajectory observable $\mathcal{O}$ is constructed upon an ensemble average that is considered with respect to the marginal distribution
\begin{eqnarray}
\langle \mathcal{O} \rangle_\alpha & = & \int \sum_{j=0}^L\mathcal{O}(\widetilde{ z}_j)  \mathrm{P}_{\mathrm{sel}}^\alpha(\widetilde{ z}_j|\zeta) 
         \mathrm{P}_{\mathrm{marg}}^\alpha(\zeta) \mathcal{D}\zeta \label{subeq:marg}. 
\end{eqnarray}
Resorting to Bayes relation~\eqref{bayes}, one checks that \eqref{subeq:marg} simplifies into the standard ensemble average \eqref{subeq:prior}. 
Owing to this statistical equivalence, information from unselected trajectories can be retrieved~\cite{AM2010,athenes:2007} provided that the sampler leaves a marginal probability distribution $\mathrm{P}_\mathrm{marg}^\theta$ invariant 
(with $\theta \neq \alpha$ in the general case). 

This requirement is met because our algorithm obeys an extended detailed balance between any consecutive joint paths $\zeta$ and ${\zeta}^\prime$.  
The fulfilled detailed balance equation writes 
\begin{equation}\label{bildettot}
\mathrm{P}_\mathrm{cond}(\mathbf{\zeta}^\prime| \widetilde{ z}_{\nu^\prime}^\prime)
\pi\left[\widetilde{ z}_{\nu^\prime}^\prime \leftarrow \widetilde{ z}_\nu   \right]
\mathrm{P}_\mathrm{sel}^\theta(\widetilde{ z}_\nu|\mathbf{\zeta})
\mathrm{P}_\mathrm{marg}^{\theta}(\zeta)
= \mathrm{P}_\mathrm{cond}(\mathbf{\zeta}| \widetilde{ z}_{\nu})
\pi\left[ \widetilde{ z}_{\nu}   \leftarrow   \widetilde{ z}_{\nu^\prime}^\prime  \right] 
\mathrm{P}_\mathrm{sel}^\theta(\widetilde{ z}_{\nu}^\prime|\widetilde{\mathbf{\zeta}}^\prime)
\mathrm{P}_\mathrm{marg}^{\theta}({\mathbf{\zeta}}^\prime), 
\end{equation} 
where $\widetilde{ z}_\nu$ and $\widetilde{ z}^\prime_{\nu^\prime}$ are the
selected trajectories belonging to joint paths $\zeta$ and ${\zeta}^\prime$
respectively (in the shifting). The quantity $\pi\left[\widetilde{
z}_{\nu^\prime}^\prime \leftarrow \widetilde{ z}_\nu   \right]$ denotes the
transition probabilities (in the shooting) from $\widetilde{ z}_\nu$ to
$\widetilde{ z}_{\nu^\prime}^\prime$ and vice versa for $\pi\left[ \widetilde{
z}_{\nu}   \leftarrow   \widetilde{ z}_{\nu^\prime}^\prime  \right]$. They obey
detailed balance with respect to $\mathrm{P}_A^\theta$ density. 
One checks that the detailed balance equation~\eqref{bildettot} is indeed satisfied by (i) resorting to the shooting balance equation, (ii) inserting Bayes relation~\eqref{bayes} within $\mathrm{P}_\mathrm{sel}^\theta( \widetilde{z}_\nu|\mathbf{\zeta})$ and $\mathrm{P}_\mathrm{sel}^\theta(\widetilde{ z}^\prime_{\nu}|\widetilde{\mathbf{\zeta}})$ and (iii) simplifying. Consequently, the distribution $\mathrm{P}_\mathrm{marg}^{\theta}$ is left invariant by our algorithm. 
 
Besides, the trajectory observable at Markov chain step \emph{m} over the
$L+1$ proposals contained in joint trajectories $\zeta^{m}$ is averaged using
the posterior probability $\mathrm{P}^\alpha_\mathrm{sel}$ of the target
ensemble  
\begin{equation}\nonumber
\overline{\mathcal{O}}^{m}_{\alpha}=\sum_{j=0}^{{L}}\mathcal{O}(\widetilde{ z}^{m}_{j}) \mathrm{P}^\alpha_\mathrm{sel} (\widetilde{ z}^m_{j}| \zeta ^ m). 
\end{equation}
The ensemble average of a trajectory observable $\mathcal{O}$ with respect to marginal distribution $P^{\alpha}_\mathrm{marg} \propto \exp \left(f_\alpha -U_\alpha \right)$ given a sample of $M$ joint trajectories distributed according to the marginal probability $P^{\theta}_\mathrm{marg} \propto \exp \left(f_\theta -U_\theta \right)$ can eventually be approximated using the following waste-recycling estimator
\begin{equation}
 \label{est_corrf_wr_unb}
{I}^\mathrm{wr}_{\alpha,\theta}\left(\mathcal{O} \right) = \sum_{m=1}^{M} 
\frac{ \overline{\mathcal{O}}^{m}_\alpha \exp\left[\hat{f}_\alpha - U_\alpha(\mathbf{\zeta}^{m}) \right] }{M \exp \left[ \hat{f}_\theta - U_\theta(\mathbf{\zeta}^{m}) \right]}
\end{equation} and again $\hat{f}_\alpha -\hat{f}_\theta$ is approximated by
solving ${I}^\mathrm{wr}_{\alpha,\theta}\left[ 1 \right]=1$. 

\subsection{Multi-state Bennett acceptance ratio}\label{MBAR}

In addition to the standard~\eqref{est_corrf_m_unb} and
waste-recycling~\eqref{est_corrf_wr_unb} estimators, we also implement the
multi-state Bennett acceptance ratio (MBAR) method developed by Shirts and
Chodera.~\cite{shirts2008statistically} 
This method is based on Bennett acceptance ratio~\cite{bennett1976efficient} and extended bridge sampling~\cite{ebs} techniques, and aims at minimizing the statistical variances associated with a series of simulations performed with $K+1$ distinct values $\{ \theta(k) \}_{0 \leq k \leq K}$ of the control parameter $\theta$. The $k$-th simulation provides one with a Markov chain consisting of $M_{\theta(k)}$ trajectories. Pooling all the data $\mathcal{O}^m$ of the observable into a single chain of size $M = \sum_{k=0}^K M_{\theta(k)}$, the waste-recycling estimate of observable $\mathcal{O}$ becomes  
\begin{equation}\label{mbar_wr}
{B}^\mathrm{wr}_{\alpha}\left(\mathcal{O} \right)=\sum_{m=1}^{{M}} 
\frac{  \overline{\mathcal{O}}^m_{\alpha} \exp{\left[ \hat{f}_{\alpha}-U_{\alpha}(\zeta^m)\right]}}{\sum_{k=0}^{K} {M}_{\theta(k)} \exp \left[ \hat{f}_{\theta(k)} -U_{\theta(k)}(\zeta^m)\right]  }. 
\end{equation}
A noticeable feature of MBAR is that the Markov chain origin of sample $z_m$ is an irrelevant information. In practice, the free energy estimates $\{ \hat{f}_{\theta(k)} \}_{1 \leq k \leq K }$ are the solutions of the set of non-linear equations 
\begin{equation} \nonumber
  {B}^\mathrm{wr}_{\theta(k)} (1) = 1 \hspace{1cm} 0 \leq k \leq K . 
\end{equation}
The normalizing constants $\hat{c}_{\theta(k)} = \exp \hat{f}_\alpha$ are evaluated up to a common multiplicative constant using the solver provided online.~\cite{chodera_site} In the waste-recycling context (Sec.~\ref{subsec:wr}), the probability masses used in MBAR are the sampled marginal probabilities $\mathrm{P}_\mathrm{marg}^{\theta(k)}(\zeta^m) \propto \exp\left[f_{\theta(k)}-U_{\theta(k)}(\zeta^m) \right] $ of~\eqref{pmarg} with $0\leq k \leq K$. 

The state-to-state time-correlation functions $C(n \tau)$ defined in Eq.~\ref{eq:correl} with $t=n \tau$ and  $n \leq L$ can easily be computed from the MBAR estimator~\eqref{mbar_wr} by setting $\alpha$ to $0$. Noticing that $\mathrm{P}^0_\mathrm{sel}(\widetilde{z}_j^m| \zeta^m)$ simplifies into $h_A(\widetilde{\mathbf{x}}_{j}^m) / \left[\sum_{j=1}^L h_A(\widetilde{\mathbf{x}}_{j}^m)\right]$ and substituting $h_B(\widetilde{\mathbf{x}}_{n+j}^m)$ for $\mathcal{O}(\widetilde{ z}_j^m)$, the conditional time-correlation along joint trajectory $\zeta^m$ reads 
\begin{equation} \label{eq:mean_reactivity}
\bar{C}_0^m (n\tau) = \frac{\sum_{j=0}^L h_A(\widetilde{\mathbf{x}}_{j}^m)h_B(\widetilde{\mathbf{x}}_{n+j}^m) }{\sum_{j=0}^L h_A(\widetilde{\mathbf{x}}_{j}^m)}. 
\end{equation}
The posterior conditional average~\eqref{eq:mean_reactivity} characterizes the reactivity of the joint trajectory and is similar to the time average usually defining time-correlation functions.~\cite{chandler1987introduction} 
Denoting ${B}^\mathrm{wr}_{0}\left( C \right)$ by $\hat{C}$, the estimator of $C(n \tau)$ reads
\begin{equation}\nonumber
\hat{C}(n \tau ) = \sum_{m=1}^{{M}}
\frac{ \bar{C}_0^m (n\tau) \exp{\left[ \hat{f}_{0}-U_{0}(\zeta^m)\right]}}{ \sum_{k=0}^{K} {M}_{\theta(k)} \exp \left[ \hat{f}_{\theta(k)} -U_{\theta(k)}(\zeta^m)\right] }.
\end{equation}
Without recycling the unselected candidates, the standard MBAR
estimator would read 
\begin{equation}\label{mbar_std}
{B}^\mathrm{std}_{\alpha}\left(\mathcal{O} \right)=\sum_{m=1}^{{M}} 
\frac{ \mathcal{O}(z^m) \exp{\left[ \hat{f}_{\alpha}-\alpha \mathcal{L}(z^m)\right]}}{\sum_{k=0}^{K} {M}_{\theta(k)} \exp \left[ \hat{f}_{\theta(k)} -\theta(k) \mathcal{L} (z^m)\right]}. 
\end{equation}
The standard MBAR form~\eqref{mbar_std} differs from the waste-recycling form~\eqref{mbar_wr} by the substitution of the probability ratios $\mathrm{P}_\mathrm{marg}^\alpha /\mathrm{P}_\mathrm{marg}^\theta = \exp\left[ f_\theta-f_\alpha- U_\theta+U_\alpha \right] $ 
by the ratios  $\mathrm{P}_A^{\alpha}/\mathrm{P}_A^{\theta}= \exp\left[ f_\theta-f_\alpha- u_\theta+u_\alpha \right] $, wherein $u_{\theta}=u_\alpha +(\theta-\alpha)\mathcal{L}$. Here, the $\hat{f}_{\theta(k)}$ estimates are solutions of $B^\mathrm{std}_{\theta(k)}(1)=1$ and are determined up to a common additive constant. 

To our knowledge, this study is the first one to implement the MBAR reweighting scheme for waste-recycling. 
However, MBAR has already been applied to nonequilibrium path-ensemble averages~\cite{minh2009optimal} for computing free energy differences or to Langevin dynamics for estimating time-correlation functions.~\cite{chodera:2011}
The path-sampling scheme of Sec.~\ref{sec:Transition-path-sampling} can be implemented with Langevin dynamics~\cite{stoltz2007path} and even with stochastic dynamics based on discrete master equations.~\cite{elmatad:2010, peters:2012} 
The implementation with Langevin dynamics involves supplementing the Verlet splitting~\eqref{eq:splitting} with Ornstein-Uhlenbeck processes on the momenta~\cite{athenes:2004,adjanor:2005,adjanor:2006,sivak:2011,leimkuhler:2012,lelievre:2012} and generating the new proposals using Stoltz's Brownian tube technique.~\cite{stoltz2007path,athenes:2008}

\section{Application to vacancy migration in $\alpha$-iron}\label{Numerical results}

SUNDAE is used to calculate the migration rate of a single vacancy in $\alpha$-Fe, 
a crystalline phase of iron with body-centered cubic (BCC) structure. Atomic
interactions of the model system are described by an embedded atom model
potential.~\cite{ackland2004} The free energy profile associated with the jump
of a neighboring atom into the vacancy 
has been calculated in Ref.~\onlinecite{AM2010} using both Monte Carlo and the harmonic approximation for temperature ranging from 
20 K to 1000 K and with a lattice parameter (average first neighbor distance)
$a=2.47 \textrm{\AA}$ corresponding to the energy minimum at $0K$. 
The free energy profile exhibits a single free energy barrier for temperatures above $T=450$ K, while for lower temperatures an intermediate metastable state appears, corresponding to the intra-site position of the jumping atom described in Ref.~\onlinecite{malerba2010comparison}. 
The overall process is clearly thermally activated for all temperatures. At 500 K, the free energy barrier calculated in Monte Carlo is $0.501 eV$. 
The reaction coordinate used to represent the migration was the projection of the atom jumping into the vacancy on an axis whose direction is parallel to the initial lattice sites of the vacancy and jumping atom (of the unrelaxed structure). 

The computational set-up in SUNDAE is as follows. 
Basin $A$ and $B$ are defined with respect to the underlying lattice whose sites
are the atomic positions of the initial structure relaxed at 0 K. 
The characteristic function $h_A$ is 1 as long as all atoms are located within a distance of $0.55 \textrm{\AA}$ from their lattice site, otherwise it is 0. 
The characteristic function $h_B$ is 1 as soon as an atom is located beyond a distance of $a/2$ from its lattice site, otherwise it is 0. Trajectories are generated with time-step $\tau=2fs$ and consists of $L=300$ steps. We have chosen the temperature of $500$ K. We use $K=51$ values of the $\theta$ parameter and set $\theta(k)=\theta_\mathrm{max}\times k/K$ with $\theta_\mathrm{max}=3.21$. 
A preliminary simulation is done to equilibrate the trajectories at $\theta = \theta_\mathrm{max}$. 
The final trajectory is active and is taken as starting point for the subsequent
equilibration runs of 500 Monte Carlo cycles performed for all $\theta(k)$
values ($0 \leq k \leq K$). 
A cycle is a shooting move followed by a shifting one.  
A few cycles are observed to inactivate the sampled trajectories at small
$\theta$ values. 
A production run consists of $2\times10^4$ Monte Carlo cycles carried out for every $\theta$ 
value independently: the sample sizes are thus $M_{\theta(k)}=2\times10^4$ for $0 \leq k \leq K$.  

\begin{figure}[ht]
 \begin{center}
 \includegraphics{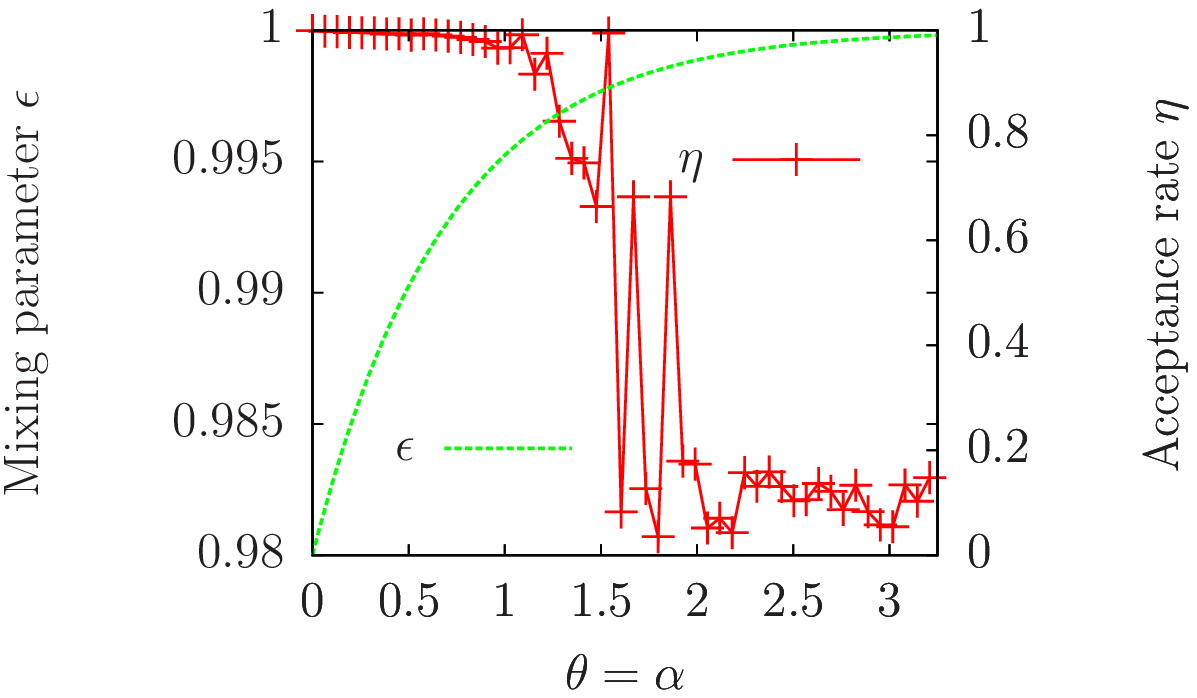}%{figure/taux_alpha.eps} 
 \caption{\label{fig:accept_stoltz} Mixing parameter 
          $\epsilon$ of Eq.~\ref{eq:stoltz} and mean acceptance rate $\eta$ 
          as a function of $\theta=\alpha$. Each $\eta$ value is calculated from 
          $2\times10^4$ shooting attempts. At large $\theta$ values, sampled paths are mainly 
          active : if $\epsilon$ is too small, then generated trial trajectories substancially 
          decorrelate, become inactive and likely fail the Metropolis test.}
\end{center} 
\end{figure}

The value of the mixing parameters $\epsilon$ (see Eq.~\ref{eq:stoltz}) is tuned
to achieve the best trade-off between acceptance (which tends to 1 in the limit
$\epsilon=1$) and decorrelation (which arises in the limit $\epsilon=0$). The
increasing function $\epsilon (\theta) = 1-2\times
100^{-1-\theta/\theta_\mathrm{max}}$ is a good compromise and ensures that the
mean acceptance rate $\eta$ does not collapse below $10\%$ with increasing
$\theta$, as shown in Fig.~\ref{fig:accept_stoltz}. 

\begin{figure}[ht]
 \centering
 \includegraphics*[scale=1.0]{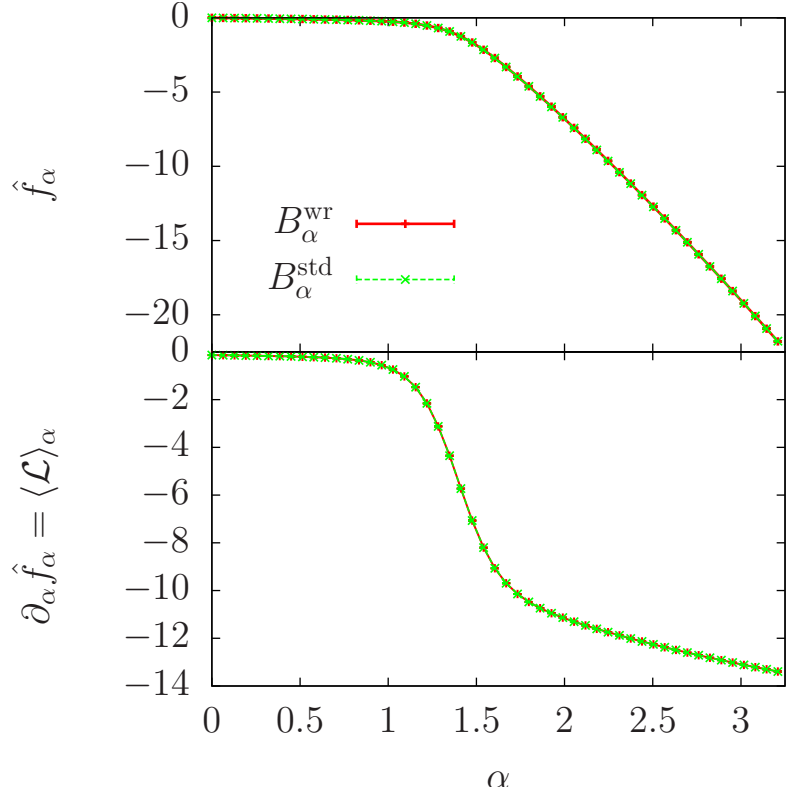}%figure/fed.eps} 
 \caption{MBAR estimates of the reduced free energy $f_\alpha$ and its 
          derivative $\partial_\alpha f_\alpha$  as a function of $\alpha$. 
          The derivative is computed from the mean activation indicator 
          $\langle \mathcal{L} \rangle_\alpha$. $B^\mathrm{wr}_\alpha$ and 
          $B^\mathrm{wr}_\alpha$ denote the waste-recycling and 
          standard MBAR estimators.}\label{fig:fed}
\end{figure}

\begin{figure}[ht]
 \centering 
 \includegraphics*[scale=1.0]{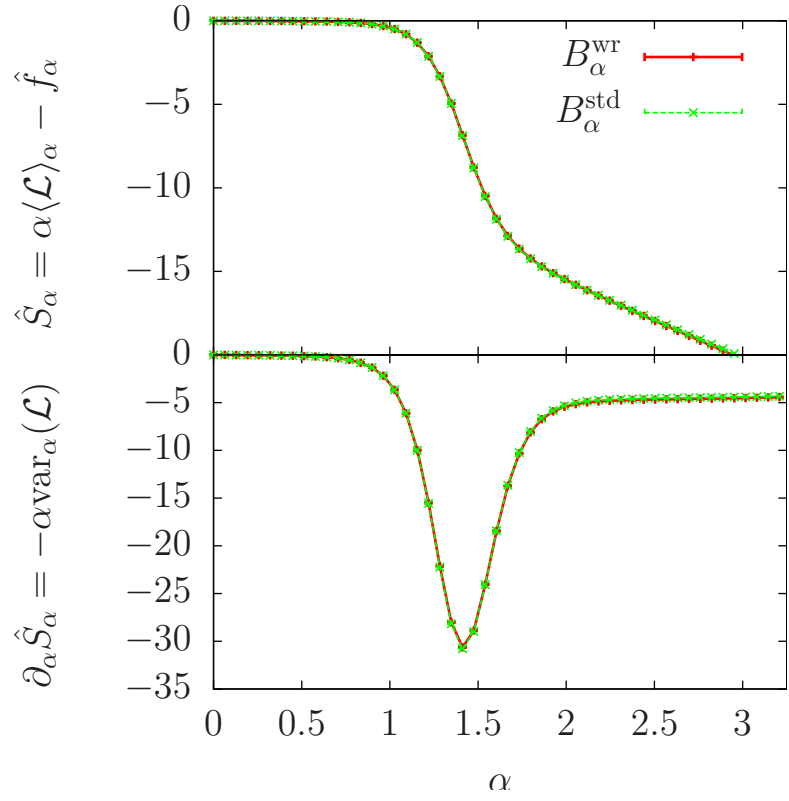}%figure/entropy.eps} 
 \caption{MBAR estimates of the entropy $S_\alpha$ and of its derivative with respect to $\alpha$. Same production run and same conditions as in Fig.~\ref{fig:fed}.}\label{fig:entropy}
\end{figure}

Various MBAR estimates have been computed as a function of $\alpha$. Free energy
estimates $\hat{f}_\alpha$ are displayed in Fig.~\ref{fig:fed} together with
their derivatives estimated from the ensemble average $\langle \mathcal{L}
\rangle_\alpha$. The entropies ${S}_\alpha$ and their derivatives, displayed in
Fig.~\ref{fig:entropy}, are estimated resorting to ensemble averages $\alpha
\langle \mathcal{L} \rangle_\alpha -f_\alpha$ and $-\alpha
\mathrm{var}_\alpha\left( \mathcal{L} \right)$, respectively. The
waste-recycling and standard forms agree with each other. 

\begin{figure}[ht]
\centering 
\includegraphics*[scale=1.0]{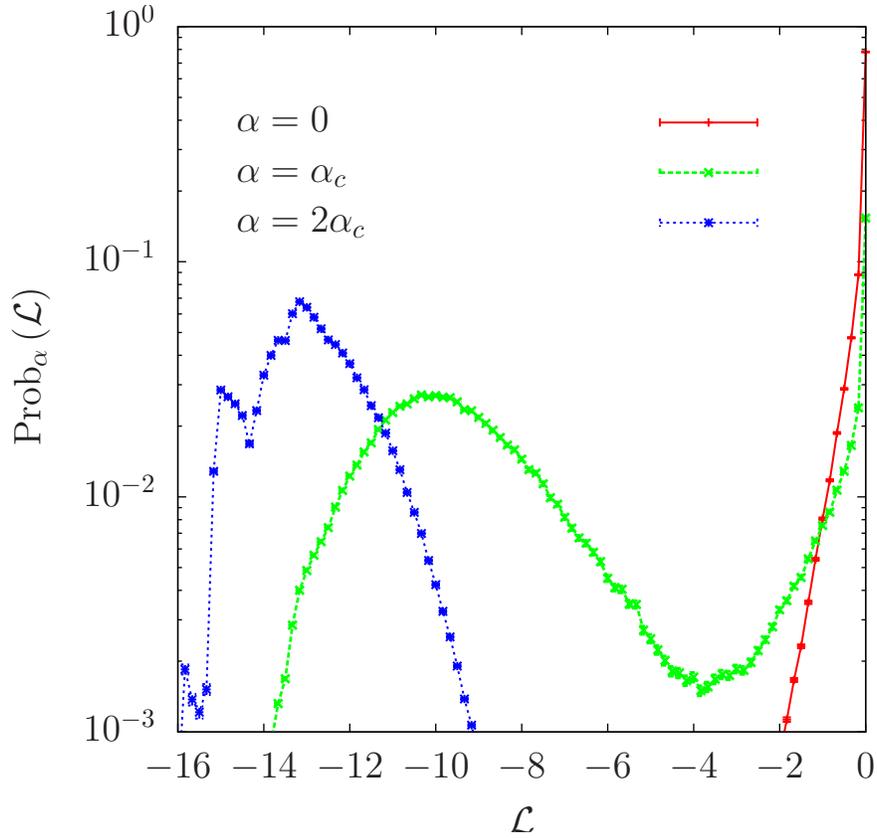}%{figure/coexistence.eps} 
\caption{The quantity $\mathrm{Prob}_\alpha (\mathcal{L})$ represents the probability to observe the value $\mathcal{L}$ during a simulation with $\theta=\alpha$. The three histograms represents the distribution of the values of the activation indicator estimated using the standard MBAR estimator at $\alpha=0$, $\alpha=\alpha_c$, and $\alpha=2\alpha_c$, where $\alpha_c=1.48$. Same production run as in Fig.~\ref{fig:fed}.}\label{fig:coexistence}
\end{figure}

The abrupt change in the slope of the reduced free energy, the inflection of the entropy and the sharp peak of the entropy derivative occurring at $\alpha_c=1.48$ are the signature of a phase transition in trajectory space.~\cite{elmatad:2010,athenes:2008} 
The first phase consists of inactive and moderately active trajectories and is referred to as ``the inactive phase''.  The
second one, consisting of active trajectories exclusively, is favored as $\alpha$
increases and is referred to as the ``active phase''. This behavior is emphasized by the histogram of the activation
indicator displayed in Fig.~\ref{fig:coexistence} for three $\alpha$ values. The
thermodynamic state is characterized by a single phase, inactive at $\alpha=0$
and active at $2\alpha_c$. In contrast at $\alpha_c$, the inactive and active
phases coexist. The phase coexistence around $\alpha_c$ results in metastability
issues in the simulations. 

\begin{figure}[ht]%[H]
\centering 
\includegraphics*[scale=1.0]{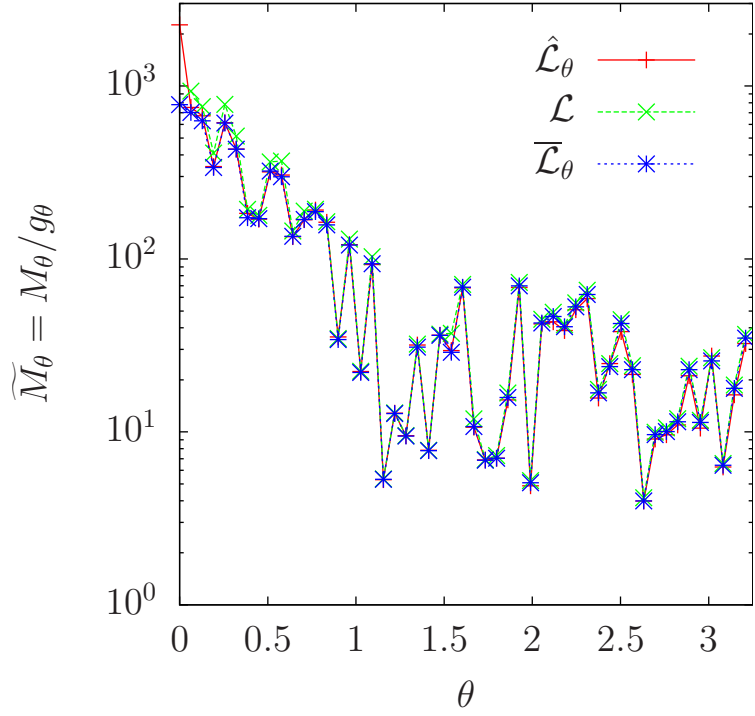}%figure/correlation_function.eps} 
\caption{Effective number of uncorrelated trajectories for all simulated values $\theta(k)$ of $\theta$ with respect to the path functionals indicated in the legend. The ratio $g_\theta=M_\theta/\widetilde{M}_\theta$ measures a statistical inefficiency ($M_\theta=2\times 10 ^4$ is the number of sampled trajectories for all the $\theta(k)$'s). }\label{fig:uncorrelated}
\end{figure}

To show that the samples are indeed correlated, we have computed the effective number of uncorrelated trajectories with respect to path observables $\mathcal{L}$, $\hat{\mathcal{L}}_\theta = \theta^{-1}(U_\theta - U_0)$ and $\bar{\mathcal{L}}_\theta$. For any observable $\mathcal{O}$ and given the sample size $M_\theta$, this quantity is defined by~\cite{shirts2008statistically}
\begin{equation}
 \widetilde{M}_\theta=M_\theta \left[ 1 + \sum_{m=1}^{M_\theta-1} \frac{M_\theta-m}{M_\theta} \times \frac{\mathrm{Cov}_\mathcal{O}(m)}{\mathrm{Cov}_\mathcal{O}(0)} \right] \label{eq:uncorrelated}, 
\end{equation}
where $\mathrm{Cov}_\mathcal{O}$ is the autocovariance function of $\mathcal{O}$ along the stationary Markov chain: 
\[ \mathrm{Cov}_\mathcal{O}(m) = \frac{1}{M_\theta-m} \sum_{p=m+1}^{M_\theta} \left[ \mathcal{O}^m \mathcal{O}^{m+p} - \mathcal{O}^m \mathcal{O}^{m+p} \right].\] The sum in~\eqref{eq:uncorrelated} corresponds to the integrated fluctuation autocorrelation time. 
Figure~\ref{fig:uncorrelated} displays the effective number of uncorrelated trajectories and clearly shows the numerical slowing-down resulting from the phase transition. 

\begin{figure}[ht]
\centering 
\includegraphics*[scale=1.0]{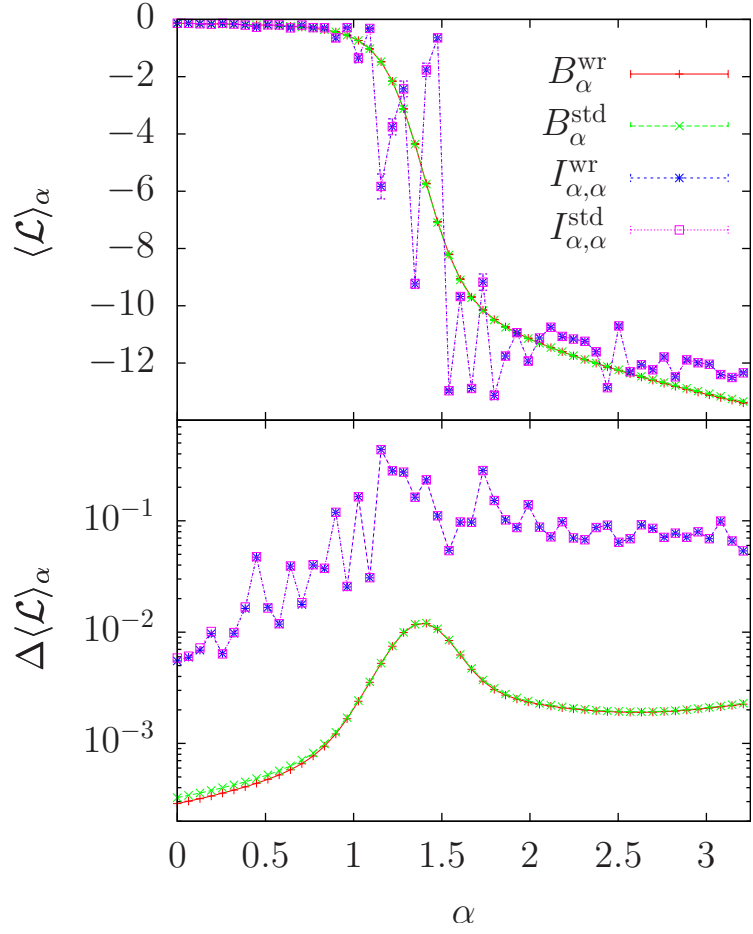}%figure/U.eps} 
\caption{Estimates of the mean activation indicator and of the standard error $\Delta \langle \mathcal{L} \rangle_\alpha$ as a function of $\alpha$ given by the MBAR estimators $B^\mathrm{wr}_{\alpha}$ and $B^\mathrm{std}_{\alpha}$ and the online estimators $I^\mathrm{wr}_{\alpha,\alpha}$ and $I^\mathrm{std}_{\alpha,\alpha}$. Same production run as in Fig.~\ref{fig:fed}. }\label{fig:U}
\end{figure}

\begin{figure}[ht]
\centering 
\includegraphics*[scale=1.0]{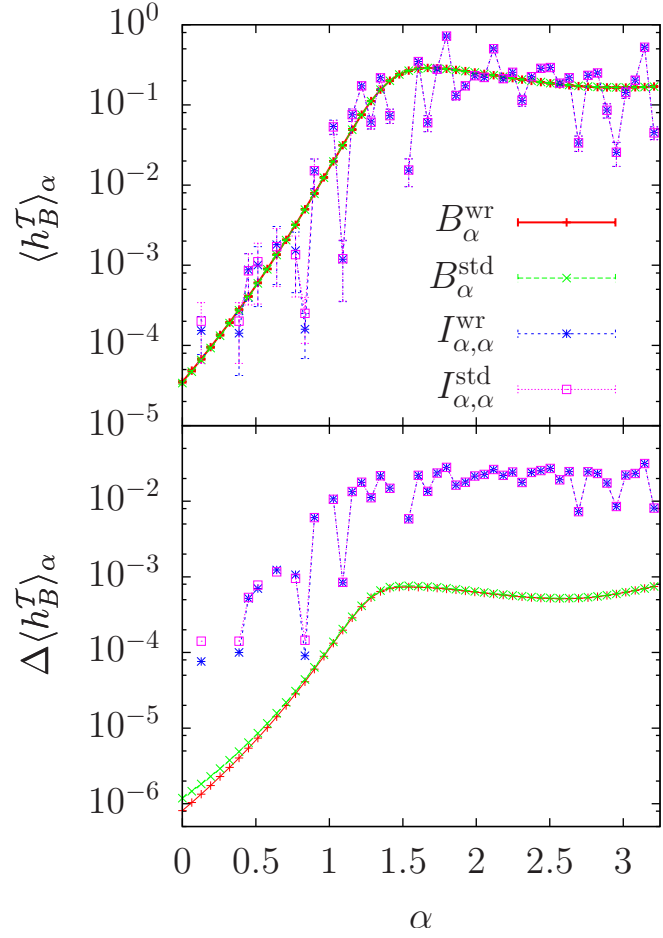}%figure/corr_alpha.eps} 
\caption{Estimates of the time-correlation function $\langle h_B^\mathcal{T} \rangle_\alpha$ and of the standard error $\Delta \langle h_B^\mathcal{T} \rangle_\alpha$ as a function of the unphysical $\alpha$ ensembles, using the MBAR estimators $B^\mathrm{wr}_{\alpha}$ and $B^\mathrm{std}_{\alpha}$ and the online estimators $I^\mathrm{wr}_{\alpha,\alpha}$ and $I^\mathrm{std}_{\alpha,\alpha}$. Same production run as in Fig.~\ref{fig:fed}.}\label{fig:reactivity_alpha}
\end{figure}

Let us now examine the standard errors for the online estimators associated with $ \langle  \mathcal{L}\rangle_\alpha$ which are displayed in Fig.~\ref{fig:U}. They were obtained by dividing each Markov chains into 200 blocks consisting of 100 consecutive points, computing the statistical variance of the 200 block estimates and eventually dividing their square roots by 200. 
Given the uncertainties, the online estimates are not reliable when $\alpha$ is close to $\alpha_c$. 
The same problematic trend holds for the estimates of the correlation observable $h_B^t(z)=h_B(\mathbf{x}(t))$ displayed in Fig.~\ref{fig:reactivity_alpha} for $t=\mathcal{T}$ as a function of $\alpha$. 

The standard error bars represented on the graphs in Figs.~\ref{fig:U} and~\ref{fig:reactivity_alpha} 
for the two MBAR estimators have been obtained using the correlated data and thus underestimate the true uncertainty in MBAR. The use of uncorrelated subsamples in MBAR (as advocated in Ref.~\onlinecite{shirts2008statistically}) was problematic due to the small subsample sizes. Nevertheless, the MBAR uncertainties allow us to compare the efficiency of waste-recycling relatively to that of waste-disposal. We only observe a substantial improvement with waste-recycling at low $\alpha$ values. At $\alpha=0$, the standard error of $B_{0}^\mathrm{wr}(h_B^\mathcal{T})$ is lower than the one of $B_{0}^\mathrm{std}(h_B^\mathcal{T})$ by almost a factor 2, amounting to a simulation speed-up of almost 4. 

\begin{figure}[ht]%[H]
\centering 
\includegraphics*[scale=1.0]{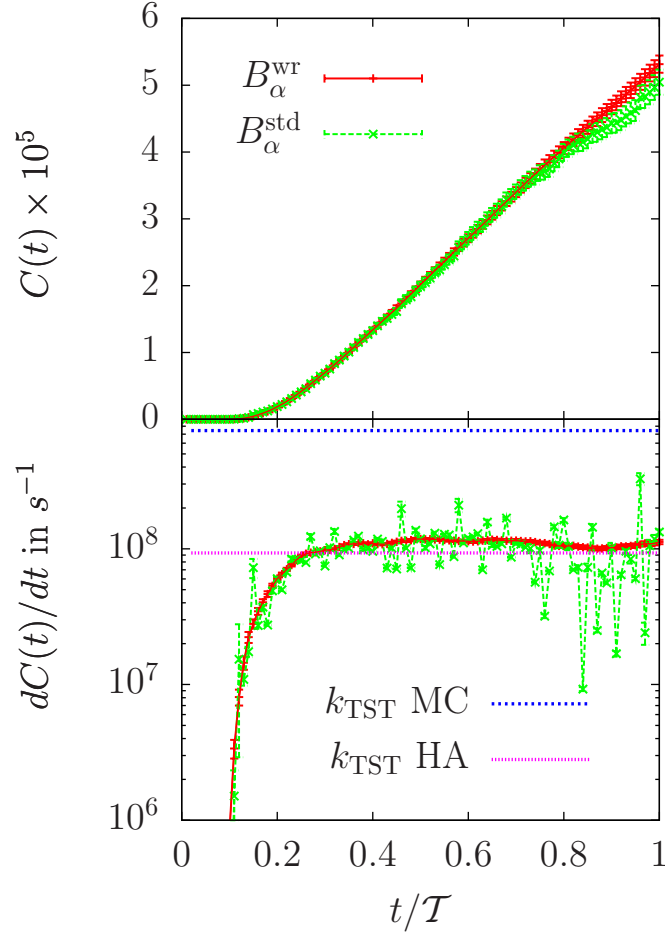}%figure/correlation_function.eps} 
\caption{Standard and waste-recycling MBAR etimates of the correlation function $C(t)=\langle h_B^\mathcal{T} \rangle_0 $ (top panel) and of its time-derivative (bottom panel). TST rates are plotted for comparison with the obtained plateau value of the time derivative.}\label{fig:plateau}
\end{figure}

Since the average $\langle h_B^t \rangle_0$ is the time-correlation function
$C(t)$, waste-recycling is useful for our problem. The information
included in the unselected reactive trajectories is relevant and improves the
accuracy of the estimates of $C(t)$. This feature is well illustrated in
Fig.~\ref{fig:plateau}: the time-correlation function grows
more linearly when estimated using the waste-recycling MBAR procedure 
than when estimated using the standard MBAR procedure. 
Concomitantly, the plateau value obtained for the derivative $dC(t)/dt$ 
is considerably less noisy with waste-recycling. 
The plateau value corresponds to the phenomenological rate constant $k_{A\rightarrow B}$. 

We have also plotted in Fig.~\ref{fig:plateau} the rates $k_\mathrm{TST} = N^c
\left(h\beta\right)^{-1}\exp\left(-\beta \Delta F_{A\rightarrow B}\right)$ given
by transition state theory~\cite{HTB1990, chandler1987introduction} for
comparison. The quantity $N^c=8$ is the coordination number of
BCC structure and $F_{A\rightarrow B}$ is the free energy barrier from $A$ to
$B$, computed by Monte Carlo in Ref.~\onlinecite{AM2010} for a cell with 1023
atoms (MC) or calculated here using the classical harmonic approximation for the
same cell of 127 atoms (HA). The small disagreements with the $k_\mathrm{TST}$
value in the HA framework may originate from anharmonicities. The bigger
disagreement with the $k_\mathrm{TST}$ value obtained by Monte Carlo may be
explained by a size effect and by the fact that the employed reaction
coordinate~\cite{AM2010} does not describe the reaction correctly, i.e. that the
transmission factor~\cite{frenkel2002understanding} associated with the reaction
coordinate is small. Unfortunately, this transmission factor was not calculated
in Ref.~\onlinecite{AM2010} and is difficult to evaluate because
the migration barrier exhibits a double hump. 

\section{Conclusion}

Motivated by the ability of eigenvalue-following~\cite{PhysRevLett.77.4358,MB1998,MW1999,HJ1999,cances2009} and Lyapunov-weighting~\cite{hinde1992chaos,amitrano1992probability,hinde1993chaotic,calvo1998chaos,W2003,TK2006,TTK2006,picciani2011simulating,geiger2010identifying} methods to locate saddle points in complex systems based on the topology of their energy surfaces, we developed and tested SUNDAE, a transition path sampling technique in which importance sampling in trajectory space does not confine the trajectory endpoints to a reactive state but rather uses the lowest eigenvalues of the Jacobian along the trajectory. 
In practice, the Lanczos algorithm is used to compute the lowest eigenvalue (as in the activation-relaxation technique~\cite{PhysRevLett.77.4358,MB1998}) and the MBAR scheme is used in combination with waste-recycling 
to unbias the dynamically activated events and to estimate the state-to-state time-correlation function. 
 
The usefulness of the approach has been demonstrated on a practical example: the migration of a vacancy in
$\alpha$-Fe. The time-correlation function associated with a vacancy jump has been computed. 
The plateau value of its time derivative has been found in good agreement with the rate 
constant given by the transition state theory. 
We also observe that the time derivative of the time-correlation function is considerably 
smoother when estimated with waste-recycling than without. 
Enabling MBAR to recycle the information contained in the unselected trajectories thus proved 
to be particularly relevant. 

To further speed-up the calculations, SUNDAE should be implemented 
in combination with a path replica exchange method~\cite{bolhuis:2008} on $\theta$,~\cite{athenes:2008} 
the conjugate parameter controlling the mean value of the path action. 
This technique is routinely used to decorrelate samples in Markov chains~\cite{athenes:re:2008} and 
should alleviate the observed numerical slowing-down 
around the phase transition induced by the use of a conjugate parameter. 
The range of $\theta(k)$ values maximizing the efficiency of replica exchange simulations will 
likely have to span this transition. 

We believe that SUNDAE will find useful applications in a wide class of fields, 
spanning from molecular biophysics to physical metallurgy,~\cite{marinica2012} 
in which the knowledge of rate constants are important input parameters in kinetic 
Monte Carlo simulations or other meso-scale models. 

The authors are grateful to John D. Chodera (University of California, Berkeley)
and Michael J. Shirts (University of Virginia) for their assistance with the
implementations of MBAR and for stimulating comments.

\bibliographystyle{elsarticle-num}

%\newpage

\end{document}